\documentclass[prd,twocolumn,nopacs,floatfix,amsmath,nofootinbib,amssymb,floatfix]{revtex4}
\usepackage{graphicx,color,dcolumn,booktabs,bm}
\usepackage{longtable,lscape}
\usepackage{txfonts}
\usepackage{overpic}
\usepackage{amssymb}
\usepackage{indentfirst}
\usepackage{feynmf}   
\usepackage{slashed}  
\usepackage{cases}
\usepackage{color}
\usepackage{multirow}
\usepackage{epstopdf}
\usepackage{longtable}
\usepackage{tikz}
\usepackage{graphicx,color,dcolumn,booktabs,bm}
\usepackage[colorlinks,
            citecolor=blue,
            anchorcolor=red,
            menucolor=red,
            linkcolor=red,
            filecolor=red,
            runcolor=red,
            urlcolor=blue,
            frenchlinks=red]{hyperref}

\graphicspath{{Figures/}} %

\allowdisplaybreaks

\begin{document}

\title{The recoil corrections, correlation functions and possible double-strange hadronic molecules}
\author{Xiao-Mei Tang$^{1}$}
\author{Li-Cheng Sheng$^{1}$}
\author{Qi Huang$^{2}$}\email{06289@njnu.edu.cn}
\author{Rui Chen$^{1,3}$}\email{chenrui@hunnu.edu.cn}

\affiliation{
$^1$Key Laboratory of Low-Dimensional Quantum Structures and Quantum Control of Ministry of Education, Department of Physics and Synergetic Innovation Center for Quantum Effects and Applications, Hunan Normal University, Changsha 410081, China\\
$^2$School of Physics and Technology, Nanjing Normal University, Nanjing 210023, China\\
$^3$Hunan Research Center of the Basic Discipline for Quantum Effects and Quantum Technologies, Hunan Normal University, Changsha 410081, China}
\date{\today}
\begin{abstract}

In this work, we perform a systematic investigation of ${K}^{(*)}{K}^{(*)}$ interactions within a one-boson-exchange model. The framework incorporates both $S-D$ wave mixing and coupled-channel effects, with effective potentials retained up to order $\mathcal{O}(1/M^2)$. By solving the coupled channel Schr\"{o}dinger equations, we can predict two double-strange molecular candidates: a $KK^*$ molecule with $I(J^P)=0(1^+)$ and a $K^*K^*$ molecule with $0(1^+)$. Our results also show that the recoil corrections play a crucial role in the formation of these two molecular candidates. Furthermore, the $S-D$ wave mixing effects contribute positively to the formation process. As a byproduct of this analysis, we extend our study to $K^{(*)}\bar{K}^{(*)}$ interactions with the same model. Our findings suggest that the $K\bar{K}^{*}$ states with $0(1^{+-}, 1^{++})$ and $K^{*}\bar{K}^{*}$ states with $0(0^{++}, 1^{+-}, 2^{++})$ can be promising molecular candidates. Additionally, we analyze the correlations between the constituent mesons, the resulting correlation functions provide additional support for our predictions of both double-strange and strangonium-like molecular states.

\end{abstract}
\pacs{***}
\maketitle

\section{introduction}

The observation of the $X(3872)$ \cite{Belle:2003nnu} and $D_{s0}(2317)$ \cite{BaBar:2003oey} in 2003 heralded a new era in hadron spectroscopy. Since then, experiments have consistently reported a series of new exotic structures, which labeled as $X/Y/Z/P_{c}/P_{cs}$ (see references~\cite{Liu:2013waa, Hosaka:2016pey, Chen:2016qju, Richard:2016eis, Lebed:2016hpi, Brambilla:2019esw, Liu:2019zoy, Chen:2022asf, Olsen:2017bmm, Guo:2017jvc, Meng:2022ozq} for more details). These observations can not only enrich the hadron spectrum, but also provides crucial opportunities to probe the non-perturbative regime of Quantum Chromodynamics (QCD). Because many structures exhibit behaviors and properties different from those predicted by the traditional quark model (mesons and baryons), theorists propose different explanations to them, such as the multiquarks, hadronic molecular states, glueballs, hybrids, and so on. 

Among these exotic configurations, the hadronic molecular state picture describes a system composed of color-singlet hadrons bound together by residual strong interactions. Notably, many of the observed $X$, $Y$, $Z$, $P_c$, and $P_{cs}$ states near relevant hadronic thresholds are frequently interpreted as prime molecular candidates \cite{Chen:2016spr,Chen:2016qju,Guo:2017jvc}. For instance, the $D_{s0}(2317)$ has been explained as a $DK$ molecular state \cite{Barnes:2003dj,Chen:2022svh,Chen:2004dy,Liu:2012zya,Yang:2021tvc,Xie:2010zza,Guo:2006fu,Guo:2006rp}, while the $X(3872)$ is often described as a $D\bar{D}^*$ molecule \cite{Swanson:2003tb,Liu:2008fh,Thomas:2008ja,Lee:2009hy,Liu:2008tn,Li:2012cs,Cai:2025inq}. The charged charmonium-like state $Z(4430)$ has been discussed as a potential $D^*\bar{D}_1$ or $D^*\bar{D}^\prime_1$ molecular candidate \cite{Liu:2007bf}. Furthermore, the $P_c(4312)$, $P_c(4440)$, and $P_c(4457)$ states are commonly attributed to molecular pentaquark configurations consisting of a charmed baryon and an anti-charmed meson \cite{Wu:2010jy,Wang:2011rga,Yang:2011wz,Uchino:2015uha,Karliner:2015ina,Chen:2019asm,Liu:2019tjn,Yamaguchi:2019seo,Chen:2019bip,Xiao:2019aya,Meng:2019ilv,PavonValderrama:2019nbk,He:2019ify,Du:2019pij,Burns:2019iih,Wang:2019ato,Xu:2025mhc,Xiao:2019mvs}.

In fact, before these observations of $X/Y/Z/P_{c}/P_{cs}$ states, the concept of molecular state had been applied to explain some hadrons. For example, in Ref. \cite{Voloshin:1976ap}, the interaction between a pair of charmed mesons was studied, which shows the possibilities of existing charmonium molecular states. Later, Rujula, Georgi, and Glashow proposed $D^*\bar{D}^*$ molecular state to explain $\psi(4040)$ \cite{De Rujula:1976qd}. By one-pion-exchange model, T\"{o}rnqvist carried out the study of deuteron-like two-meson bound states ($D\bar{D}^*$ and $D^*\bar{D}^*$) \cite{Tornqvist:1993ng,Tornqvist:1993vu}. While $f_0(980)$ and $a_0(980)$ as $K\bar{K}$ molecular states was proposed by Weinstein and Isgur \cite{Weinstein:1982gc,Weinstein:1983gd,Weinstein:1990gu}. Owing to this rich history and its continued relevance in interpreting new phenomena, the study of hadronic molecular states has matured into a vital and highly active subfield within hadron physics.

Recently, the LHCb Collaboration observed an narrow state in the $D^0D^0\pi^+$ mass spectrum, it locates just below the $D^{*+}D^0$ mass threshold \cite{LHCb:2021vvq,LHCb:2021auc}. Its mass with respect to the $D^0D^{*+}$ threshold and width are
\begin{eqnarray}
\delta m &=& m_{T_{cc}^+}-(m_{D^0}+m_{D^{*+}})= -273\pm 61\pm 5^{+11}_{-14}~\text{keV},\nonumber\\
\Gamma &=& 410\pm 165\pm 43^{+18}_{-38}~\text{keV},\nonumber
\end{eqnarray}
respectively. According to the final states, this state contains two charm quarks, an anti-$u$, and an anti-$d$ quark at least, and the spin-parity is $J^P=1^+$. Due to the near threshold character, this doubly charmed tetraquark $T_{cc}^+$ is very likely to be the $S-$wave $DD^*$ molecular state with a very small binding energy \cite{Chen:2021cfl,Dong:2021bvy,Feijoo:2021ppq,Albaladejo:2021vln,Fleming:2021wmk,Meng:2021jnw,Du:2021zzh,Lin:2022wmj,Cheng:2022qcm,Ke:2021rxd,Ling:2021bir,Liu:2019yye,Yan:2021wdl,Jin:2021cxj,Xin:2021wcr,Shi:2022slq,Ortega:2022efc,Du:2023hlu,Wang:2022jop,Chen:2023fgl,Chen:2021vhg}. Actually, before the observation, authors in Ref. \cite{Li:2012ss,Xu:2017tsr,Li:2021zbw,Janc:2004qn,Ohkoda:2012hv} have already predicted the existence of the $DD^*$ molecule with $I(J^P)=0(1^+)$, and the predicted mass is very consistent with the mass of the newly $T_{cc}^+$. 

Obviously, searching for the double strange molecular counterpart of the $T_{cc}$ state can provide a important test of molecular state picture. Especially, it can give indirect test of molecular state picture for the $T_{cc}^+$ state. Very recent, Wang \textit{et. al.} studied the spectroscopic properties of the $\bar{K}^{(*)}\bar{K}^{(*)}$ molecules by using the one-boson-exchange (OBE) model, and they found the $\bar{K}\bar{K}^*$ state with $I(J^P)=0(1^+)$ and the $\bar{K}^*\bar{K}^*$ state with $0(1^+)$ can be the most likely double-strangeness molecular candidates \cite{Wang:2024kke}. In their work, the effective Lagrangians were constructed in the heavy meson formalism, which inherently neglects recoil corrections associated with the three-momenta of the interacting particles. However, in Refs.~\cite{Zhao:2014gqa,Zhao:2015mga}, the authors kept the OBE effective potentials for the $D^{(*)}\bar{D}^{(*)}$ and $B^{(*)}\bar{B}^{(*)}$ interactions up to the order of $1/M^2$ with $M$ being the mass of the interaction particles, and they found that the higher recoil corrections can turn out to be important for the very loosely bound molecular states. 

Given these considerations, it is very interesting to investigate the $K^{(*)}K^{(*)}$ interactions with recoil corrections, in particular, the higher recoil corrections can be non-ignorable due to the smaller mass of the strange mesons. In the following, we still derive the OBE effective potentials first, which include the interactions from the exchange of the light pseudoscalar $(\pi,\eta)$, vector $(\rho,\omega,\phi)$ and scalar $\sigma$ mesons. The effective Lagrangians used are constructed in an $SU(3)$ symmetry. We keep the OBE effective potentials up to the order $1/M^2$. Then, we search for possible double strange molecular tetraquarks by solving the coupled channel Schr\"{o}dinger equations. 

Meanwhile, we calculate the correlation functions between the two strange mesons of the possible hadronic molecular states, which can be another way called as femtoscopic technique to detect a two-body molecular system in experiment, apart from the traditional amplitude analysis. For example, by measuring the correlation function $C(k)$, the ALICE collaboration successfully obtained a very similar correlation function as lattice results \cite{ALICE:2020mfd,Morita:2016auo,HALQCD:2018qyu,HALQCD:2019wsz}, where a $p-\Omega^-$ bound state and a relatively strong attraction between $p$ and $\Xi^-$ are predicted, and such conclusions are also verified by the STAR collaboration very recently \cite{STAR:report}. In addition, as a byproduct, we also discuss the existence of the possible $K^{(*)}\bar{K}^{(*)}$ molecules by adopting the same approach.

This paper is organized as follow. After introduction, the detailed deduction of the effective potential of the $K^{(*)}K^{(*)}$ systems is given in Sec. \ref{sec2}. In Sec. \ref{sec3}, we present the numerical results. Finally, the paper ends with a summary in Sec. \ref{sec4}.

\section{The $K^{(*)}K^{(*)}$ interactions}\label{sec2}

\begin{figure}[!htbp]
\center
\includegraphics[width=2.5in]{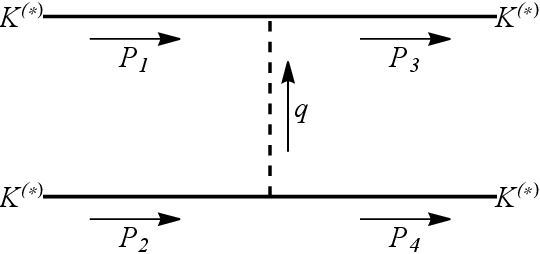}\\
\caption{Feynman diagram of the $K^{(*)}K^{(*)}\to K^{(*)}K^{(*)}$ process.}\label{fig1}
\end{figure}

In this section, we deduce the OBE effective potentials for the $K^{(*)}K^{(*)}$ systems. In Figure \ref{fig1}, we present the corresponding Feynman diagram. Here, $P_1(E_1,\vec{p})$, $P_2(E_2,-\vec{p})$, $P_3(E_3,\vec{p}^{\prime})$, and $P_4(E_4,-\vec{p}^{\prime})$ stand for four momentum of the initial and final particles, respectively.  And we define $q=(E_3-E_1,\vec{p}^{\prime}-\vec{p})=(E_2-E_4,\vec{p}^{\prime}-\vec{p})$ and $\vec{k}=(\vec{p}+\vec{p}^{\prime})$.

The relevant Lagrangians can be constructed as \cite{Lin:1999ad,Nagahiro:2008mn,Chen:2019uvv} 
\begin{eqnarray}
\mathcal{L}_{\sigma} &=& g_{\sigma }m_K\bar{K} K\sigma-g_{\sigma }m_{K^*}\bar{K}^{*}\cdot K^{*}\sigma, \label{lag1}\\
\mathcal{L}_{PPV} &=& \frac{ig}{2\sqrt{2}}\langle\partial^{\mu}P\left(PV_{\mu}-V_{\mu}P\right)\rangle,\label{lag1}\\
\mathcal{L}_{VVP} &=& \frac{g_{VVP}}{\sqrt{2}}\epsilon^{\mu\nu\alpha\beta}
\left\langle\partial_{\mu}V_{\nu}\partial_{\alpha}V_{\beta}P\right\rangle,\\
\mathcal{L}_{VVV}  &=& \frac{ig}{2\sqrt{2}}\langle\partial^{\mu}V^{\nu}
               \left(V_{\mu}V_{\nu}-V_{\nu}V_{\mu}\right)\rangle. \label{lag4}
\end{eqnarray}
Here, $P$ and $V$ respectively stand for the pseudoscalar and vector mesons in SU(3) octet, i.e.,
\begin{eqnarray}
P &=& \left(\begin{array}{ccc}
\frac{\pi^0}{\sqrt{2}}+\frac{\eta}{\sqrt{6}} &\pi^+ &K^+ \nonumber\\
\pi^- &-\frac{\pi^0}{\sqrt{2}}+\frac{\eta}{\sqrt{6}} &K^0 \nonumber\\
K^- &\bar{K}^0 &-\sqrt{\frac{2}{3}}\eta
\end{array}\right),\label{p}\\
{V} &=& \left(\begin{array}{ccc}
\frac{\rho^0}{\sqrt{2}}+\frac{\omega}{\sqrt{2}}  &\rho^+      &K^{*+}\\
\rho^- &-\frac{\rho^0}{\sqrt{2}}+\frac{\omega}{\sqrt{2}}      &K^{*0}\\
K^{*-}      &\bar{K}^{*0}    &\phi
\end{array}\right).\label{p}
\end{eqnarray}
Coupling constants in above Lagrangian are  estimated as \cite{Liu:2011xc, Chen:2017xat, Kaymakcalan:1983qq},  
$g_{\sigma} = -3.65$, $g=12.00$, $g_{vvp} = \frac{3g^2}{32\sqrt{2}\pi^2 f_{\pi}}$, $f_{\pi} = 132$ MeV. 

Upon expanding the Lagrangians in Eqs. (\ref{lag1})-(\ref{lag4}) in terms of the SU(3) octet pseudoscalar and vector meson fields, we obtain the following explicit expressions:
\begin{eqnarray}
\mathcal{L}_{\pi,\eta} &=& \frac{ig}{4}\left[\left(\bar{K}^{*\mu} K-\bar{K} K^{*\mu}\right)\left(\bm{\tau}\cdot\partial_{\mu}\bm{\pi}+\frac{\partial_{\mu}{\eta}}{\sqrt{3}}\right)\right.\nonumber\\
   &&\left.+\left(\partial_{\mu}\bar{K} K^{*\mu}-\bar{K}^{*\mu}\partial_{\mu}K\right)\left(\bm{\tau}\cdot\bm{\pi}+\frac{\eta}{\sqrt{3}}\right)\right]\nonumber\\
   &&+g_{VVP}\varepsilon_{\mu\nu\alpha\beta}  \partial^{\mu}\bar{K}^{*\nu}\partial^{\alpha}K^{*\beta}\left(\bm{\tau}\cdot\bm{\pi}+\frac{\eta}{\sqrt{3}}\right) ,\\
\mathcal{L}_{\rho,\omega,\phi} &=& \frac{ig}{4}\left[\bar{K}\partial_{\mu}K-\partial_{\mu}\bar{K}K\right]\left(\bm{\tau}\cdot\bm{\rho}^{\mu}+{\omega}^{\mu}-\sqrt{2}\phi^{\mu}\right)\nonumber\\
       &&+\frac{ig}{4} \left[\left(\bar{K}_{\mu}^*\partial^{\mu}K^{*\nu}-\partial^{\mu}\bar{K}^{*\nu} K_{\mu}^*\right)\left(\bm{\tau}\cdot\bm{\rho}_{\nu}+\omega_{\nu}-\sqrt{2}\phi_{\nu}\right)\right.\nonumber\\
&&\left.+\left(\partial^{\mu}\bar{K}^{*\nu}K_{\nu}^*-\bar{K}_{\nu}^*\partial^{\mu}K^{*\nu}\right)
       \left(\bm{\tau}\cdot\bm{\rho}_{\mu}+\omega_{\mu}-\sqrt{2}\phi_{\mu}\right)\right.\nonumber\\
       &&\left.+\left(\bar{K}_{\nu}^* K^*_{\mu}-\bar{K}_{\mu}^*K^*_{\nu}\right)\left(\bm{\tau}\cdot\partial^{\mu}\bm{\rho}^{\nu}+\partial^{\mu}\omega^{\nu}-\sqrt{2}\partial^{\mu}\phi^{\nu}\right)\right]\nonumber\\
       &&+g_{VVP}\varepsilon_{\mu\nu\alpha\beta} \left(\partial^{\mu}\bar{K}^{*\nu}K+\bar{K}\partial^{\mu}{K}^{*\nu}\right)\nonumber\\
&&\left(\bm{\tau}\cdot\partial^{\alpha}\bm{\rho}^{\beta}+\partial^{\alpha}{\omega}^{\beta}-\sqrt{2}\partial^{\alpha}\phi^{\beta}\right).
\end{eqnarray}

With these Lagrangians, we then adopt the effective Lagrangians approach to write the amplitudes for $K^{(*)}K^{(*)}\to K^{(*)}K^{(*)}$ processes. In order to obtain all the momentum-related terms, we make a lorentz boost to the polarization vectors of the $K^*$ vector mesons, $\epsilon_{\lambda}=(0,\vec{\epsilon}_{\lambda})$, i.e.,
\begin{eqnarray}
\epsilon_{\lambda}(p) &=& \left(\frac{\vec{p}\cdot\vec{\epsilon}_{\lambda}}{m}, \vec{\epsilon}_{\lambda}+\frac{\vec{p}\cdot\vec{\epsilon}_{\lambda}}{m(p^0+m)}\vec{p}\right).
\end{eqnarray}
Here, $m$ and $p=(p^0,\vec{p})$ correspond to the mass and the four momentum of the vector mesons in the laboratory frame, respectively. 

In a Breit approximation, we can then obtain the effective potentials in the momentum space, i.e., $\mathcal{V}(\vec{p}',\vec{p}) = -\mathcal{M}(\vec{p}',\vec{p})/\sqrt{\prod_i2M_i\prod_f2M_f}$. Here, $\mathcal{M}(\vec{p}',\vec{p})$ stands for the scattering amplitude. $M_i$ and $M_f$ are the masses of the initial and final states, respectively. In Table \ref{potential}, we present the OBE effective potentials for the $K^{(*)}K^{(*)}\to K^{(*)}K^{(*)}$ processes in the momentum space. By performing the Fourier transformation, we can finally obtain the effective potential in the coordinate space, as shown in Table \ref{transformation}, i.e.,
\begin{eqnarray}
\mathcal{V}(\vec{r}',\vec{r}) = \int \frac{d^3\vec{k}d^3\vec{q}}{(2\pi)^{3}}e^{i\vec{k}\cdot(\vec{r}'-\vec{r})+i\frac{\vec{q}}{2}\cdot(\vec{r}+\vec{r})}\mathcal{V}(\vec{p}',\vec{p})\mathcal{F}^2(q^2,m_E^2).
\end{eqnarray}
Here, we introduce a form factor in every interaction vertex, which can compensate the off-shell effects of the exchanged bosons. And we take the monopole type form factor, $\mathcal{F}(q^2,m_E^2)=(\Lambda^2-m_E^2)/(\Lambda^2-q^2)$. Here, $\Lambda$, $m_E$, and $q$ are the cutoff, the mass and four momentum of the exchanged bosons, respectively. According to the experience of the nucleon-nucleon interactions \cite{Tornqvist:1993ng,Tornqvist:1993vu}, the reasonable cutoff value is taken around 1.00 GeV.

\begin{table*}[!htbp]
\renewcommand\tabcolsep{0.2cm}
\renewcommand{\arraystretch}{1.7}
\caption{A summary of the OBE effective potentials for the $K^{(*)}K^{(*)}\to K^{(*)}K^{(*)}$ processes in the momentum space.} \label{potential}
\begin{tabular}{cccl}
\toprule[1pt]\toprule[1pt]
Processes   &Diagram &Exchanged mesons &OBE effective potentials \\\hline
$KK\to KK$  &\multirow{3}{*}{\begin{tikzpicture}
    \fill[gray!30] (-0.5,-0.75) rectangle (0.5,0.75);
    \draw[line width=0.5mm] (-1,0.5) -- (1,0.5);
    \draw[line width=0.5mm] (-1,-0.5) -- (1,-0.5);
    \draw[dashed] (0,-0.5) -- (0,0.5);
\end{tikzpicture}} 
&$\sigma$   &$V_{\sigma}^{a}= -\frac{g_{\sigma}^2}{4}\frac{1}{\vec{q}^2+m_{\sigma}^2}$\\
&& $\rho,\omega,\phi$  &$V_{V}^{a}=-\frac{g^2}{16}\frac{1}{\vec{q}^2+m_{V}^2}-\frac{g^2}{16}\frac{\vec{k}^2}{m_{K}^2}\frac{1}{\vec{q}^2+m_{V}^2}$\\\cline{3-4}
&&Total    &{$V^{I=1}=V_{\sigma}^a+V_{\rho}^a+V_{\omega}^a+2V_{\phi}^a$}\\\hline
$KK^*\to KK^*$  &\multirow{3}{*}{\begin{tikzpicture}
    \fill[gray!30] (-0.5,-0.75) rectangle (0.5,0.75);
    \draw[line width=0.5mm] (-1,0.5) -- (1,0.5);
    \draw[line width=1.0mm] (-1,-0.5) -- (1,-0.5);
    \draw[dashed] (0,-0.5) -- (0,0.5);
\end{tikzpicture}} 
&$\sigma$   &$V_{\sigma}^{b}=-\frac{g_{\sigma}^2}{4}\frac{(\vec{\epsilon}_4^{\dag} \cdot \vec{\epsilon}_2)}{\vec{q}^2+m_{\sigma}^2}-\frac{g_{\sigma}^2}{8m_{K^*}^2}\frac{(\vec{q} \cdot \vec{\epsilon}_4^\dag) (\vec{q} \cdot \vec{\epsilon}_2)}{\vec{q}^2+m_{\sigma}^2}-\frac{g_{\sigma}^2}{8m_{K^*}^2}\frac{ (\vec{q} \times \vec{k}) \cdot(\vec{\epsilon}_2 \times \vec{\epsilon}_4^\dag)}{\vec{q}^2+m_{\sigma}^2}$\\
&& $\rho,\omega,\phi$  &$V_{V}^{b}=\frac{g^2}{16}\frac{(\vec{\epsilon}_4^{\dag} \cdot \vec{\epsilon}_2)}{\vec{q}^2+m_{V}^2}+\frac{g^2}{16}\frac{\vec{k}^2}{m_{K}m_{K^*}}\frac{(\vec{\epsilon}_4^{\dag} \cdot \vec{\epsilon}_2)}{\vec{q}^2+m_{V}^2}-\frac{g^2}{16}(\frac{1}{2m_{K^*}^2}+\frac{1}{m_{K}m_{K^*}})\frac{ (\vec{q} \times \vec{k})\cdot(\vec{\epsilon}_2 \times \vec{\epsilon}_4^\dag)}{\vec{q}^2+m_{V}^2}$\\
&&&\quad\quad\quad$-\frac{g^2}{16}\frac{1}{2m_{K^*}^2}\frac{ (\vec{q} \cdot \vec{\epsilon}_4^\dag)   (\vec{q} \cdot \vec{\epsilon}_2)}{\vec{q}^2+m_{V}^2}$\\\cline{3-4}
 &\multirow{3}{*}{\begin{tikzpicture}
    \fill[gray!30] (-0.5,-0.75) rectangle (0.5,0.75);
    \draw[line width=0.5mm] (-1,0.5) -- (0,0.5);
    \draw[line width=1.0mm] (-1,-0.5) -- (0,-0.5);
    \draw[line width=0.5mm] (0,-0.5) --  (1,-0.5);
    \draw[line width=1.0mm] (0,0.5)  --  (1,0.5);
    \draw[dashed] (0,-0.5) -- (0,0.5);
\end{tikzpicture}}
&$\pi$  &$V_{\pi}^{c}=-\frac{g^2}{16}\frac{(m_{K}+m_{K^*})^2}{4m_{K}m_{K^*}^3}\frac{(\vec{q} \cdot \vec{\epsilon}_3^\dag)    (\vec{q} \cdot \vec{\epsilon}_2)}{\vec{q}^2-m_{\pi_0}^2}+\frac{g^2}{16}\frac{m_{K}^2-m_{K^*}^2}{2m_{K}m_{K^*}^3}\frac{(\vec{k} \times \vec{q})\cdot(\vec{\epsilon}_2 \times \vec{\epsilon}_3)}{\vec{q}^2-m_{\pi_0}^2}+\frac{g^2}{16}\frac{(m_{K}-m_{K^*})^2}{m_{K}m_{K^*}^3}\frac{(\vec{k} \cdot \vec{\epsilon}_3^\dag)     (\vec{k} \cdot \vec{\epsilon}_2)}{\vec{q}^2-m_{\pi_0}^2}$\\
&&$\eta$  &$V_{\eta}^{c}=-\frac{g^2}{16}\frac{(m_{K}+m_{K^*})^2}{4m_{K}m_{K^*}^3}\frac{(\vec{q} \cdot \vec{\epsilon}_3^\dag)  (\vec{q} \cdot \vec{\epsilon}_2)}{\vec{q}^2+m_{\eta_0}^2}+\frac{g^2}{16}\frac{m_{K}^2-m_{K^*}^2}{2m_{K}m_{K^*}^3}\frac{(\vec{k} \times \vec{q})\cdot(\vec{\epsilon}_2 \times \vec{\epsilon}_3)}{\vec{q}^2+m_{\eta_0}^2}+\frac{g^2}{16}\frac{(m_{K}-m_{K^*})^2}{m_{K}m_{K^*}^3}\frac{(\vec{k} \cdot \vec{\epsilon}_3^\dag) (\vec{k} \cdot \vec{\epsilon}_2)}{\vec{q}^2+m_{\eta_0}^2}$\\
&& $\rho,\omega,\phi$  &$V_{V}^{c}=-\frac{g_{vvp}^2}{16}\frac{(m_{K}+m_{K^*})^2}{m_{K}m_{K^*}}\frac{(\vec{\epsilon}_2\times\vec{q})\cdot(\vec{\epsilon}_3^\dag \times\vec{q})}{\vec{q}^2+m_{V_0}^2}+\frac{g_{vvp}^2}{4}\frac{(m_{K}-m_{K^*})^2}{m_{K}m_{K^*}}\frac{(\vec{\epsilon}_2\times\vec{k})\cdot(\vec{\epsilon}_3^\dag \times\vec{k})}{\vec{q}^2+m_{V_0}^2}$\\
&&&\quad\quad\quad$+\frac{g_{vvp}^2}{8}(\frac{m_{K}}{m_{K^*}}-\frac{m_{K^*}}{m_{K}})\frac{(\vec{k} \times \vec{q})\cdot(\vec{\epsilon}_2 \times \vec{\epsilon}_3^\dag)}{\vec{q}^2+m_{V_0}^2}$\\\cline{3-4}
&&Total    &{$V^{I=0}=V_{\sigma}^b-3V_{\rho}^b+V_{\omega}^b+2V_{\phi}^b-\left(-3V_{\pi}^c+\frac{1}{3}V_{\eta}^c-3V_{\rho}^c+V_{\omega}^c+2V_{\phi}^c\right)$}\\
    &&&{$V^{I=1}=V_{\sigma}^b+V_{\rho}^b+V_{\omega}^b+2V_{\phi}^b+\left(V_{\pi}^c+\frac{1}{3}V_{\eta}^c+V_{\rho}^c+V_{\omega}^c+2V_{\phi}^c\right)$}\\\hline
$K^*K^*\to K^*K^*$  &\multirow{3}{*}{\begin{tikzpicture}
    \fill[gray!30] (-0.5,-0.75) rectangle (0.5,0.75);
    \draw[line width=1.0mm] (-1,0.5) -- (1,0.5);
    \draw[line width=1.0mm] (-1,-0.5) -- (1,-0.5);
    \draw[dashed] (0,-0.5) -- (0,0.5);
\end{tikzpicture}}
&$\sigma$  &$V_{\sigma}^d=-\frac{g_\sigma^2}{4}\frac{(\vec{\epsilon}_3^\dag \cdot \vec{\epsilon}_1) (\vec{\epsilon}_4^\dag \cdot \vec{\epsilon}_2)}{\vec{q}^2+m_{\sigma}^2}-\frac{g_\sigma^2}{8m_{K^*}^2}\frac{(\vec{\epsilon}_3^\dag \cdot \vec{\epsilon}_1) (\vec{\epsilon}_4^\dag \cdot \vec{q}) (\vec{\epsilon}_2 \cdot \vec{q})}{\vec{q}^2+m_{\sigma}^2}+\frac{g_\sigma^2}{8m_{K^*}^2}\frac{(\vec{\epsilon}_3^\dag \cdot \vec{\epsilon}_1)  (\vec{\epsilon}_4^\dag \times \vec{\epsilon}_2) \cdot (\vec{k} \times \vec{q})}{\vec{q}^2+m_{\sigma}^2}$\\
&&&\quad\quad\quad$-\frac{g_\sigma^2}{8m_{K^*}^2}\frac{(\vec{\epsilon}_4^\dag \cdot \vec{\epsilon}_2) (\vec{\epsilon}_3^\dag \cdot \vec{q}) (\vec{\epsilon}_1 \cdot \vec{q}) }{\vec{q}^2+m_{\sigma}^2}+\frac{g_\sigma^2}{8m_{K^*}^2}\frac{(\vec{\epsilon}_4^\dag \cdot \vec{\epsilon}_2)  (\vec{\epsilon}_3^\dag \times \vec{\epsilon}_1) \cdot (\vec{k} \times \vec{q})}{\vec{q}^2+m_{\sigma}^2}$\\
&&$\pi,\eta$   &{$V_{P}^d=\frac{g_{vvp}^2}{4}\frac{[\vec{q}\cdot (\vec{\epsilon}_1\times\vec{\epsilon}_3^\dag)]  [\vec{q}\cdot (\vec{\epsilon}_2\times\vec{\epsilon}_4^\dag)]}{\vec{q}^2+m_{P}^2}$}\\
&&$\rho,\omega,\phi$  &$V_{V}^d=\frac{g^2}{16}\frac{(\vec{\epsilon}_2 \cdot \vec{\epsilon}_4^\dag) (\vec{\epsilon}_1 \cdot\vec{\epsilon}_3^\dag)}{\vec{q}^2+m_{V}^2}+\frac{g^2}{16}\frac{\vec{k}^2}{m_{K^*}^2}\frac{(\vec{\epsilon}_2 \cdot \vec{\epsilon}_4^\dag) (\vec{\epsilon}_1 \cdot \vec{\epsilon}_3^\dag)}{\vec{q}^2+m_{V}^2}+\frac{g^2}{16}\frac{3}{2m_{K^*}^2}\frac{(\vec{\epsilon}_2 \cdot \vec{\epsilon}_4^\dag)  (\vec{k} \times \vec{q}) \cdot (\vec{\epsilon}_1 \times \vec{\epsilon}_3^\dag)}{\vec{q}^2+m_{V}^2}$\\
&&&\quad\quad\quad$+\frac{g^2}{16}\frac{3}{2m_{K^*}^2}\frac{(\vec{\epsilon}_1 \cdot \vec{\epsilon}_3^\dag)  (\vec{k} \times \vec{q}) \cdot (\vec{\epsilon}_2 \times \vec{\epsilon}_4^\dag) }{\vec{q}^2+m_{V}^2}+\frac{g^2}{16}\frac{1}{m_{K^*}^2}\frac{[(\vec{\epsilon}_1 \times \vec{\epsilon}_3^\dag) \times \vec{q}] \cdot [(\vec{\epsilon}_2 \times \vec{\epsilon}_4^\dag) \times \vec{q}]}{\vec{q}^2+m_{V}^2}$\\
&&&\quad\quad\quad$-\frac{g^2}{16}\frac{1}{2m_{K^*}^2}\frac{ (\vec{\epsilon}_2 \cdot \vec{\epsilon}_4^\dag) (\vec{q} \cdot \vec{\epsilon}_1) (\vec{q} \cdot \vec{\epsilon}_3^\dag) }{\vec{q}^2+m_{V}^2}-\frac{g^2}{16}\frac{1}{2m_{K^*}^2}\frac{ (\vec{\epsilon}_1 \cdot \vec{\epsilon}_3^\dag) (\vec{q} \cdot \vec{\epsilon}_2) (\vec{q} \cdot \vec{\epsilon}_4^\dag) }{\vec{q}^2+m_{V}^2}$\\\cline{3-4}
&&Total    &$V^{I=0}=V_{\sigma}^d-3V_{\rho}^d+V_{\omega}^d+2V_{\phi}^d-3V_{\pi}^d+\frac{1}{3}V_{\eta}^d$\\
    &&&$V^{I=1}=V_{\sigma}^d+V_{\rho}^c+V_{\omega}^d+2V_{\phi}^d+V_{\pi}^d+\frac{1}{3}V_{\eta}^d$\\\hline
$KK^*\to K^*K^*$  &\multirow{3}{*}{\begin{tikzpicture}
    \fill[gray!30] (-0.5,-0.75) rectangle (0.5,0.75);
    \draw[line width=0.5mm] (-1,0.5) -- (0,0.5);
    \draw[line width=1.0mm] (0,0.5) -- (1,0.5);
    \draw[line width=1.0mm] (-1,-0.5) -- (1,-0.5);
    \draw[dashed] (0,-0.5) -- (0,0.5);
\end{tikzpicture}}
&$\pi,\eta$   &$V_{P}^e=\frac{g_{vvp}g}{4}\frac{q_1-m_{K}+m_{K^*}}{4m_{K}^{1/2}m_{K^*}^{3/2}}\frac{i (\vec{k} \cdot \vec{\epsilon}_3^\dag)  [\vec{q} \cdot (\vec{\epsilon}_4^\dag \times \vec{\epsilon}_2)]}{\vec{q}^2+m_{P_1}^2}+\frac{g_{vvp}g}{4}\frac{q_1-m_{K}-3m_{K^*}}{4m_{K}^{1/2}m_{K^*}^{3/2}}\frac{i (\vec{q} \cdot \vec{\epsilon}_3^\dag)  [\vec{q} \cdot (\vec{\epsilon}_4^\dag \times \vec{\epsilon}_2)]}{\vec{q}^2+m_{P_1}^2}$\\
&&$\rho,\omega,\phi$  &$V_{V}^e=-\frac{g_{vvp}g}{4}\frac{q_1-4m_{K^*}}{4m_{K}^{1/2}m_{K^*}^{3/2}}\frac{i(\vec{\epsilon}_4^\dag \cdot \vec{\epsilon}_2)  [ \vec{\epsilon}_3^\dag \cdot (\vec{k} \times \vec{q}) ]}{\vec{q}^2+m_{V_1}^2}+\frac{g_{vvp}g}{4}\frac{q_1^2}{2m_{K}^{1/2}m_{K^*}^{5/2}}\frac{i(\vec{k} \cdot \vec{\epsilon}_4^\dag) [ \vec{k} \cdot (\vec{\epsilon}_2 \times \vec{\epsilon}_3^\dag) ] }{\vec{q}^2+m_{V_1}^2}$\\
 &&&\quad\quad\quad$
 +\frac{g_{vvp}g}{4}\frac{q_1-m_{K^*}}{2m_{K}^{1/2}m_{K^*}^{3/2}}\frac{i (\vec{q} \cdot \vec{\epsilon}_4^\dag)  [ (\vec{\epsilon}_2 \times \vec{\epsilon}_3^\dag) \cdot \vec{q} ]}{\vec{q}^2+m_{V_1}^2}+\frac{g_{vvp}g}{4}\frac{4m_{K^*}q_1-q_1^2}{8m_{K}^{1/2}m_{K^*}^{5/2}}\frac{i [\vec{\epsilon}_2 \times (\vec{\epsilon}_4^\dag \times \vec{\epsilon}_3^\dag) ] \cdot (\vec{k} \times \vec{q})}{\vec{q}^2+m_{V_1}^2}$\\
 &&&\quad\quad\quad$-\frac{g_{vvp}g}{4}\frac{q_1}{2m_{K}^{1/2}m_{K^*}^{3/2}}\frac{i [\vec{\epsilon}_2 \cdot (\vec{\epsilon}_4^\dag \times \vec{\epsilon}_3^\dag) ]  (\vec{k} \cdot \vec{q})}{\vec{q}^2+m_{V_1}^2}$\\\cline{3-4}
   &\multirow{3}{*}{\begin{tikzpicture}
    \fill[gray!30] (-0.5,-0.75) rectangle (0.5,0.75);
    \draw[line width=1.0mm] (-1,0.5) -- (0,0.5);
    \draw[line width=1.0mm] (0,0.5) -- (1,0.5);
    \draw[line width=0.5mm] (-1,-0.5) -- (0,-0.5);
   \draw[line width=1.0mm] (0,-0.5) -- (1,-0.5);
    \draw[dashed] (0,-0.5) -- (0,0.5);
\end{tikzpicture}} &$\pi,\eta$   &$V_{P}^f=\frac{g_{vvp}g}{4}\frac{q_2-m_{K}+m_{K^*}}{4m_{K}^{1/2}m_{K^*}^{3/2}}\frac{i (\vec{k} \cdot \vec{\epsilon}_3^\dag)  [\vec{q} \cdot (\vec{\epsilon}_4^\dag \times \vec{\epsilon}_1)]}{\vec{q}^2+m_{P_2}^2}+\frac{g_{vvp}g}{4}\frac{q_2-m_{K}-3m_{K^*}}{4m_{K}^{1/2}m_{K^*}^{3/2}}\frac{i (\vec{q} \cdot \vec{\epsilon}_3^\dag)  [\vec{q} \cdot (\vec{\epsilon}_4^\dag \times \vec{\epsilon}_1)]}{\vec{q}^2+m_{P_2}^2}$\\
&&$\rho,\omega,\phi$ &$V_{V}^f=-\frac{g_{vvp}g}{4}\frac{q_2-4m_{K^*}}{4m_{K}^{1/2}m_{K^*}^{3/2}}\frac{i(\vec{\epsilon}_4^\dag \cdot \vec{\epsilon}_1)  [ \vec{\epsilon}_3^\dag \cdot (\vec{k} \times \vec{q}) ]}{\vec{q}^2+m_{V_2}^2}+\frac{g_{vvp}g}{4}\frac{q_2^2}{2m_{K}^{1/2}m_{K^*}^{5/2}}\frac{i(\vec{k} \cdot \vec{\epsilon}_4^\dag) [ \vec{k} \cdot (\vec{\epsilon}_1 \times \vec{\epsilon}_3^\dag) ] }{\vec{q}^2+m_{V_2}^2}$\\
 &&&\quad\quad\quad$
 +\frac{g_{vvp}g}{4}\frac{q_2-m_{K^*}}{2m_{K}^{1/2}m_{K^*}^{3/2}}\frac{i (\vec{q} \cdot \vec{\epsilon}_4^\dag)  [ (\vec{\epsilon}_1 \times \vec{\epsilon}_3^\dag) \cdot \vec{q} ]}{\vec{q}^2+m_{V_2}^2}+\frac{g_{vvp}g}{4}\frac{4m_{K^*}q_2-q_2^2}{8m_{K}^{1/2}m_{K^*}^{5/2}}\frac{i [\vec{\epsilon}_1 \times (\vec{\epsilon}_4^\dag \times \vec{\epsilon}_3^\dag) ] \cdot (\vec{k} \times \vec{q})}{\vec{q}^2+m_{V_2}^2}$\\
&&&\quad\quad\quad$-\frac{g_{vvp}g}{4}\frac{q_2}{2m_{K}^{1/2}m_{K^*}^{3/2}}\frac{i [\vec{\epsilon}_1 \cdot (\vec{\epsilon}_4^\dag \times \vec{\epsilon}_3^\dag) ]  (\vec{k} \cdot \vec{q})}{\vec{q}^2+m_{V_2}^2}$\\\cline{3-4}
&&Total    &$V^{I=0}=-\frac{3}{\sqrt{2}}V_{\rho}^e+\frac{1}{\sqrt{2}}V_{\omega}^e+\sqrt{2}V_{\phi}^e-\frac{3}{\sqrt{2}}V_{\pi}^e+\frac{1}{3\sqrt{2}}V_{\eta}^e$\\
&&&\quad\quad\quad$-\left(-\frac{3}{\sqrt{2}}V_{\rho}^f+\frac{1}{\sqrt{2}}V_{\omega}^f+\sqrt{2}V_{\phi}^f-\frac{3}{\sqrt{2}}V_{\pi}^f+\frac{1}{3\sqrt{2}}V_{\eta}^f \right)$\\
         &&&$V^{I=1}=\frac{1}{\sqrt{2}}V_{\rho}^e+\frac{1}{\sqrt{2}}V_{\omega}^e+\sqrt{2}V_{\phi}^e+\frac{1}{\sqrt{2}}V_{\pi}^e+\frac{1}{3\sqrt{2}}V_{\eta}^e$\\
         &&&\quad\quad\quad$+\left(\frac{1}{\sqrt{2}}V_{\rho}^f+\frac{1}{\sqrt{2}}V_{\omega}^f+\sqrt{2}V_{\phi}^f+\frac{1}{\sqrt{2}}V_{\pi}^f+\frac{1}{3\sqrt{2}}V_{\eta}^f \right)$\\
\bottomrule[1pt]\bottomrule[1pt]
\end{tabular}
\end{table*}

\begin{table}[!htbp]
\renewcommand\tabcolsep{0.15cm}
\renewcommand{\arraystretch}{1.8}
\caption{A summary of the Fourier transformation for all the OBE effective potentials. Here, we define two useful functions, i.e., $S(\hat{r}, \vec{A}, \vec{B}) = 3 (\hat{r} \cdot \vec{A})(\hat{r} \cdot \vec{B}) - \vec{A} \cdot \vec{B}$ and $Y(\Lambda,m,r) = \frac{1}{4\pi r}(e^{-mr}-e^{-\Lambda r}) - \frac{\Lambda^2 - m^2}{8\pi\Lambda}e^{-\Lambda r}$.} \label{transformation}
\begin{tabular}{c|l}
\toprule[1pt]\toprule[1pt]
$V(\bm{q},\bm{k})$     &$V(r)$\\\hline
$\frac{1}{\vec{q}^2 + m^2}$    &$Y(\Lambda,m,r)$\\
$\frac{\vec{q}^2}{\vec{q}^2 + m^2}$   &$-\nabla^2 Y(\Lambda,m,r)$\\
$\frac{\vec{k}^2}{\vec{q}^2 + m^2}$    &$\frac{1}{4}\nabla^2 Y(\Lambda, m, r)-\frac{1}{2}\left\{\nabla^2, Y(\Lambda, m, r)\right\}$\\
$\frac{\vec{k} \cdot \vec{q}}{q^2 + m^2}$  &$- \frac{1}{2}\nabla^2 Y(\Lambda,m,r)-\nabla Y(\Lambda,m,r) \cdot \nabla$\\
$\frac{i\vec{A} \cdot (\vec{q} \times \vec{k})}{\vec{q}^2 + m^2}$   &$ \frac{1}{r} \frac{\partial}{\partial r}  Y(\Lambda, m, r)\vec{A} \cdot \vec{L}$\\
$\frac{(\vec{A} \cdot \vec{q})(\vec{B} \cdot \vec{q})}{\vec{q}^2 + m^2}$  &$-\frac{\vec{A} \cdot \vec{B}}{3}\nabla^2 Y(\Lambda, m, r)
-\frac{1}{3}S(\hat{r}, \vec{A}, \vec{B})r\frac{\partial}{\partial r}\frac{1}{r}\frac{\partial}{\partial r}Y(\Lambda, m, r)$\\
$\frac{(\vec{A} \cdot \vec{k})(\vec{B} \cdot \vec{k})}{\vec{q}^2 + m^2}$   &$-\frac{\vec{A} \cdot\vec{B}}{12}\nabla^2 Y(\Lambda,m,r) -\frac{S(\hat{r},\vec{A},\vec{B})}{12}r\frac{\partial}{\partial r}\frac{1}{r}\frac{\partial}{\partial r}Y(\Lambda, m, r)$\\
   &$ -\frac{1}{3}\left((\vec{A}\cdot\vec{B})+S(\hat{r},\vec{A},\vec{B})\right)Y(\Lambda,m,r)\nabla^2$\\
   &$ -\frac{1}{3}\left((\vec{A}\cdot\vec{B})+S(\hat{r},\vec{A},\vec{B})\right) \nabla Y(\Lambda,m,r) \cdot \nabla$\\
$\frac{(\vec{A}\cdot\vec{k})(\vec{B}\cdot\vec{q})}{q^{2}+m^{2}}$   &$-\frac{1}{6}\left((\vec{A}\cdot\vec{B})\nabla^{2}Y(\Lambda,m,r)+S(\hat{r},\vec{A},\vec{B})r\frac{\partial}{\partial r}\frac{1}{r}\frac{\partial}{\partial r}Y(\Lambda,m,r)\right)$\\
   &$-\frac{1}{3}\left(S(\hat{r},\vec{A},\vec{B})+\vec{A}\cdot\vec{B}\right)\nabla Y(\Lambda,m,r) \cdot \nabla$\\
$\frac{(\vec{A} \times \vec{q})\cdot(\vec{B} \times \vec{q})}{\vec{q}^2 + m^2}$    &$-\frac{2 \vec{A} \cdot \vec{B}}{3}\nabla^2 Y(\Lambda, m, r)  + \frac{S(\hat{r}, \vec{A}, \vec{B})}{3}r\frac{\partial}{\partial r}\frac{1}{r}\frac{\partial}{\partial r}Y(\Lambda, m, r)$\\
$\frac{(\vec{A} \times \vec{k})\cdot(\vec{B} \times \vec{k})}{\vec{q}^2 + m^2}$   &$-\frac{\vec{A} \cdot\vec{B}}{6}\nabla^2 Y(\Lambda,m,r) +\frac{S(\hat{r},\vec{A},\vec{B})}{12}r\frac{\partial}{\partial r}\frac{1}{r}\frac{\partial}{\partial r}Y(\Lambda, m, r) $\\
  &$+\left(-\frac{2}{3}(\vec{A}\cdot\vec{B})+\frac{1}{3}S(\hat{r},\vec{A},\vec{B})\right)Y(\Lambda,m,r)\nabla^2 $\\
  &$\left(-\frac{2}{3}(\vec{A}\cdot\vec{B})+\frac{1}{3}S(\hat{r},\vec{A},\vec{B})\right) \nabla Y(\Lambda,m,r) \cdot \nabla.$\\
\bottomrule[1pt]\bottomrule[1pt]
\end{tabular}
\end{table}

In this work, we consider both the $S-D$ wave mixing effects and the coupled channel effects. Here, we first construct the wave functions for the investigated systems, which include the flavor wave function, the spatial wave function, and the spin-orbit wave functions. The flavor wave functions can be constructed as
\begin{eqnarray}
    \left|0,0\right\rangle &=&\frac{1}{\sqrt{2}}\left(K^{(*)+}K^{(*)0}-K^{(*)0}K^{(*)+}\right),\\
    \left|1,1\right\rangle&=& K^{(*)+}K^{(*)+},\\
    \left|1,-1\right\rangle&=& K^{(*)0}K^{(*)0},\\
    \left|1,0\right\rangle &=&\frac{1}{\sqrt{2}}\left(K^{(*)+}K^{(*)0}+K^{(*)0}K^{(*)+}\right),
\end{eqnarray}
for the $KK$ and $K^*K^*$ systems, and
\begin{eqnarray}
|0,0\rangle &=& \frac{1}{2}\left(K^+K^{*0}-K^{*+}K^0-K^0K^{*+}+K^{*0}K^+\right),\\
|1,1\rangle &= &\frac{1}{\sqrt{2}}\left(K^+K^{*+}+K^{*+}K^+\right),\\
|1,-1\rangle &= &\frac{1}{\sqrt{2}}\left(K^0K^{*0}+K^{*0}K^0\right),\\
|1,0\rangle &=& \frac{1}{2}\left(K^+K^{*0}+K^{*+}K^0+K^0K^{*+}+K^{*0}K^+\right), 
\end{eqnarray}
for the $KK^*$ systems, respectively. The general expressions for the spin-orbit wave functions for the $K^{(*)}K^{(*)}$ systems can be expresses as
\begin{eqnarray}
\left|K{K}^*({}^{2S+1}L_{J})\right\rangle &=& \sum_{m_L,m_S}C^{JM}_{1m,Lm_L}\epsilon^{m_{S}}_{\lambda}Y_{L,m_L},\\
\left|K^*{K}^*({}^{2S+1}L_{J})\right\rangle &=& \sum_{m,m',m_L} C^{JM}_{Sm_S,Lm_L}C^{Sm_S}_{1m,1m'}\epsilon^{m'}_{\lambda'}\epsilon^{m}_{\lambda}Y_{Lm_L}.\quad
\end{eqnarray}
Here, $C^{JM}_{1m,Lm_L}$, $C^{JM}_{Sm_S,Lm_L}$, and $C^{Sm_S}_{1m,1m'}$ are the Clebsch-Gordan coefficients, the $Y_{Lm_L}$ is the spherical harmonics function. $\vec{\epsilon_{\lambda}}$ stands for the polarization vector, which has the form of $\epsilon^m_{\pm}=\mp\frac{1}{\sqrt{2}}(\epsilon^m_x\pm i\epsilon^m_y)$ and $\epsilon^m_0=\epsilon^m_z$. When sandwiching the operators summarized in Table \ref{transformation} by the discussed spin-orbit wave functions, one can obtain a serial of non-zero elements matrices, as shown in Table \ref{Matrix elements}.

\begin{table}[!htbp]
\renewcommand\tabcolsep{0.02cm}
\renewcommand{\arraystretch}{1.8}
\caption{Matrix elements $\langle \mathcal{O}_i\rangle$ for the spin-spin, spin-orbit, and tensor force interactions operators in the OBE effective potentials. Here, $\langle \vec{\epsilon}_4^{\dag} \cdot \vec{\epsilon}_2\rangle=\langle\vec{\epsilon}_3^{\dag} \cdot \vec{\epsilon}_2\rangle=\langle( \vec{\epsilon}_3^{\dag}\cdot\vec{\epsilon}_1) ( \vec{\epsilon}_4^{\dag} \cdot \vec{\epsilon}_2)\rangle=\langle i\vec{\epsilon}_4^{\dag} \cdot (\vec{\epsilon}_2\times \vec{\epsilon}_3^{\dag})\rangle/\sqrt{2}=\mathcal{I}$, with $\mathcal{I}$ being the unit matrix. $\langle i\vec{L}\cdot(\vec{\epsilon}_2 \times \vec{\epsilon}_3^{\dag})\rangle_{J=1}=\text{diag}(0,3)$.}  \label{Matrix elements}
 \centering
 \begin{tabular}{clll}
\toprule[1pt]\toprule[1pt]
&$S(\hat{r},\vec{\epsilon}_4^{\dag},\vec{\epsilon}_2)$ &$i\vec{L}\cdot(\vec{\epsilon}_2\times\vec{\epsilon}_4^{\dag})$
&$S(\hat{r},\vec{\epsilon}_3^{\dag},\vec{\epsilon}_2)$ \\ \hline
$J=1$\quad  &$\begin{pmatrix}0 & -\sqrt{2}\\ -\sqrt{2}& 1\end{pmatrix} $
&$\begin{pmatrix}0 & 0\\ 0& 3\end{pmatrix}$
 &$\begin{pmatrix}0 & -\sqrt{2}\\ -\sqrt{2}& 1\end{pmatrix}$
\\ \midrule[1pt]
	& $S(\hat{r},\vec{\epsilon}_4^{\dag},(\vec{\epsilon}_2\times \vec{\epsilon}_3^{\dag} ))$ &$(\vec{\epsilon}_4^{\dag} \cdot \vec{\epsilon}_2)(\vec{\epsilon}_3^{\dag}\cdot\vec{L}) $
 & $\vec{\epsilon}_2 \times (\vec{\epsilon}_4^{\dag}\times \vec{\epsilon}_3^{\dag})\times \vec{L}$  
 \\ \hline
$J=1$ &$\begin{pmatrix}0 & \sqrt{\frac{5}{3}}\\1& -\sqrt{\frac{5}{6}}\end{pmatrix}$
&$\begin{pmatrix}0 &0\\ 0& \frac{3\sqrt{2}}{5}\end{pmatrix} $
&$\begin{pmatrix}0 & 0\\ 0& -3\sqrt{2}\end{pmatrix}$ \\ \midrule[1pt]
& $(\vec{\epsilon}_3^{\dag}\cdot\vec{\epsilon}_1) S(\hat{r},\vec{\epsilon}_4^{\dag},\vec{\epsilon}_2)$ &$(\vec{\epsilon}_3^{\dag}\cdot\vec{\epsilon}_1)(i\vec{L}\cdot(\vec{\epsilon}_2 \times \vec{\epsilon}_4^{\dag}))$
	& $S(\hat{r},\vec{\epsilon}_1\times\vec{\epsilon}_3^{\dag},\vec{\epsilon}_2\times\vec{\epsilon}_4^{\dag})$
 \\ \hline
$J=0$  &$\begin{pmatrix}0&-\sqrt{2}\\ -\sqrt{2}& 1\end{pmatrix}$ &$\begin{pmatrix}0 & 0\\ 0& -3\end{pmatrix}$ 
&$\begin{pmatrix}0 & \sqrt{2}\\ \sqrt{2}& 2\end{pmatrix}$  
\\ 
$J=1$ &$\begin{pmatrix}0&\frac{1}{\sqrt{2}}\\\frac{1}{\sqrt{2}}& -\frac{1}{2}\end{pmatrix}$ &$\begin{pmatrix}0 & 0\\ 0& -\frac{3}{2}\end{pmatrix}$ 
&$\begin{pmatrix}0 & -\sqrt{2}\\ -\sqrt{2}& 1\end{pmatrix}$
\\ 
$J=2$ 
 &{\scriptsize{$\begin{pmatrix}0&-\sqrt{\frac{2}{5}}&-\sqrt{\frac{7}{10}}\\ -\sqrt{\frac{2}{5}}& 0&\frac{2}{\sqrt{7}}\\-\sqrt{\frac{7}{10}}& \frac{2}{\sqrt{7}}&-\frac{3}{14}\end{pmatrix}$}} &$\begin{pmatrix}0&0&0\\0&0&0\\0&0&-\frac{3}{2}\end{pmatrix}$
&{\scriptsize{$\begin{pmatrix}0 &\sqrt{\frac{2}{5}}&-\sqrt{\frac{14}{5}}\\ \sqrt{\frac{2}{5}}& 0&-\frac{2}{\sqrt{7}}\\-\sqrt{\frac{14}{5}}&-\frac{2}{\sqrt{7}}& -\frac{3}{7}\end{pmatrix}$ }}  
\\ \midrule[1pt]
&$( \vec{\epsilon}_3^{\dag}\times\vec{\epsilon}_1)\cdot( \vec{\epsilon}_4^{\dag} \times \vec{\epsilon}_2)$ 
 
&$(\vec{\epsilon}_4^{\dag}\cdot\vec{\epsilon}_2)(i\vec{L}\cdot(\vec{\epsilon}_1 \times \vec{\epsilon}_3^{\dag}))$  & $(\vec{\epsilon}_4^{\dag}\cdot\vec{\epsilon}_2)S(\hat{r},\vec{\epsilon}_3^{\dag},\vec{\epsilon}_1)$
	
 \\ \hline
$J=0$
&$\begin{pmatrix}2 & 0\\ 0& -1\end{pmatrix}$

&$\begin{pmatrix}0 & 0\\ 0& -3\end{pmatrix}$
 &$\begin{pmatrix}0&-\sqrt{2}\\-\sqrt{2}& 1\end{pmatrix}$
 \\ 
$J=1$
&$\begin{pmatrix}1 & 0\\ 0& 1\end{pmatrix} $
&$\begin{pmatrix}0 & 0\\ 0& -\frac{3}{2}\end{pmatrix}$    
&$\begin{pmatrix}0&\frac{1}{\sqrt{2}}\\\frac{1}{\sqrt{2}}& \frac{-1}{2}\end{pmatrix}$
\\ 
$J=2$
  &$\begin{pmatrix}-1 &0&0\\0&2&0\\0&0&-1 \end{pmatrix}$
&$\begin{pmatrix}0&0&0\\ 0&0&0\\0&0&-\frac{3}{2}\end{pmatrix}$
 &{\scriptsize{$\begin{pmatrix}0&-\sqrt{\frac{2}{5}}&-\sqrt{\frac{7}{10}}\\ -\sqrt{\frac{2}{5}}& 0&\frac{2}{\sqrt{7}}\\-\sqrt{\frac{7}{10}}& \frac{2}{\sqrt{7}}&-\frac{3}{14}\end{pmatrix}$ }} 
\\ \bottomrule[1pt]\bottomrule[1pt]
\end{tabular}  
 \end{table}

Apart from the spectra calculation, in this work, we also adopt the femtoscopic technique to measure the correlation between the strange mesons, which can be measured in experiments as follows \cite{Liu:2024uxn,STAR:2014dcy,STAR:2018uho,ALICE:2020mfd}
\begin{eqnarray}
    C(E_p)=\xi(E_p) \frac{N_{same}(E_p)}{N_{mixed}(E_p)},
\end{eqnarray}
Here, $E_p \equiv \frac{\hbar^2\boldsymbol{q}^2}{2\mu}$ is the relative energy between two particles, with $\mu$ and $\boldsymbol{q}$ being the two-body reduced mass and relative momentum, respectively. $\xi(E_p)$ denotes the corrections for experimental effects, $N_{same}(E_p)$ is the events where two particles are produced in the same collision with given $\boldsymbol{q}$, and $N_{mixed}(E_p)$ is
the events where two particles are produced in different collisions. 

In the theoretical side, the correlation between two particles can be calculated through the Koonin-Pratt (KP) formula as \cite{Koonin:1977fh,Pratt:1990zq,Bauer:1992ffu}
\begin{eqnarray}
	C(E_p) = \int \mathrm{d} \boldsymbol{r} S_{12}(r)|\Psi(\boldsymbol{r}, \boldsymbol{q})|^{2},
\end{eqnarray}
where $\boldsymbol{q}$ is still the relative momentum between two particles in the center-of-mass, $\boldsymbol{r}$ is the relative coordinate, $\Psi(\boldsymbol{r}, \boldsymbol{q})$ is the relative scattering wave function obtained by solving the coupled-channel Schr\"{o}dinger equation, and $S_{12}(r)$ is the normalized pair source function represented by the relative coordinate as
\begin{eqnarray}
    S_{12}(r) = \frac{1}{(4 \pi R^2)^{3/2}} \text{exp}(-\frac{r^2}{4R^2}).
\end{eqnarray}
Here, $R$ is the size parameter of the source, which is a phenomenological parameter that usually related with the characteristics of the experimental detector. In theoretical calculation, to match the usual size of the molecule state, in this work, we set $R\approx 1$ fm as Refs.~\cite{Yan:2024aap,Yan:2025hpa} did.

In general, if the potential between two particles is relatively strong enough to bind them together into a loosely bound state or a virtual state or a shape-type resonance, as both experimental and theoretical results show \cite{Liu:2024uxn,STAR:2014dcy,STAR:2018uho,ALICE:2020mfd,Morita:2016auo,HALQCD:2018qyu,HALQCD:2019wsz}, the correlation between two particles will be significantly larger than 1 at low relative energy, this phenomena is easy to understand since for a loosely bound state, at low relative energy, the two constituent particles are still tend to form into a pair, which makes them much more easier to be observed simultaneously in the same collision than different collisions, i.e., $N_{same}(E_p) \gg N_{mixed}(E_p)$ when $E_p \to \boldsymbol{0}$. In the same way, if a near-threshold Feshbach-type resonance appears, the closer to that threshold (or energy region), the more chance that the two constituent particles will show up together. As a result, for near-threshold Feshbach-type resonance, there will generate a cusp-like structure in the correlation function $C(E_p)$, as Ref.~\cite{Liu:2024uxn} shows.

\section{numerical result}\label{sec3}

After deduced the OBE effective potentials, we next solve the coupled channel Schr\"{o}dinger equations to search for the bound state solutions. We vary the cutoff in the range of $\Lambda\leq2.00$ GeV as its reasonable range is around 1.00 GeV due to the experience of deuteron \cite{Tornqvist:1993ng,Tornqvist:1993vu}. For a loosely bound molecular state, the binding energy is from several to several tens MeV, and the root-mean-square (RMS) radius $r_{RMS}$ is around 1.00 fm or much larger. In the following, we analyze our numerical results by adopting these two bound state properties.

\subsection{The $K^{(*)}K^{(*)}$ systems}

We first investigate the potential for a double-strange molecular state predominantly composed of the $KK$ system. For the single $KK({}^1S_0)$ channel with $I(J^P)=1(0^+)$, the allowed interactions arise from the exchange of the scalar $\sigma$ meson and the vector $(\rho, \omega, \phi)$ mesons, as detailed in Table \ref{potential}. The $\sigma$-exchange potential provides an attractive contribution, while the $\rho$, $\omega$, and $\phi$ exchanges are repulsive. In the cutoff range of $\Lambda\leq2.00$ GeV, we cannot obtain the loosely bound state solutions. This conclusion remains unchanged even after the inclusion of $S$-$D$ wave mixing, coupled-channel effects (with the $KK({}^1S_0)$, $K^*K^*({}^1S_0)$, and $K^*K^*({}^5D_0)$ channels), recoil corrections, and spin-orbit forces. Consequently, we rule out the existence of molecular candidates predominantly composed of $KK$ systems with quantum numbers $I(J^P)=1(0^+)$ The current results are also consistent with the conclusions in Ref. \cite{Wang:2024kke}.

\begin{table}[!htbp]
\renewcommand\tabcolsep{0.2cm}
\renewcommand{\arraystretch}{1.7}
\caption{The bound state solutions for the $KK^{*}$ systems with $I(J^P)=0(1^+)$. Here, case I and case II present the results without and with the momentum-related terms, respectively. $E$ and $r_{\text{RMS}}$ stand for the binding energy and the the root-mean-square radius, respectively. The unites for the cutoff $\Lambda$, $E$, and $r_{RMS}$ are GeV, MeV, and fm, respectively.} \label{num1}
\begin{tabular}{cccccccc}
\toprule[1pt]\toprule[1pt]
    &\multicolumn{3}{c}{Case I}  &   &\multicolumn{3}{c}{Case II}\\\cline{2-4}\cline{6-8}
    &$\Lambda$    &$E$     &$r_{\text{RMS}}$ &
           &$\Lambda$    &$E$     &$r_{\text{RMS}}$\\\midrule[1pt]
$S-$wave    
&1.36  &$-0.39$  &10.51   &    &1.36   &$-5.45$   & 2.51\\
&1.42  &$-4.44$   &2.80   &    &1.42  &$-18.29$   & 1.46 \\
&1.48  &$-12.32$   &1.80  &    &1.48  &$-38.24$   &1.03\\\hline
$S-D$   
       &1.36  &$-1.17$   &5.37 &   &1.36   &$-6.50$    &2.33  \\
 wave  &1.42  &$-6.14$   &2.43  &  &1.42   &$-19.64$    &1.40\\
       &1.48  &$-14.51$  &1.66 &   &1.48   &$-39.66$    &1.01\\\hline
Coupled   &1.36    &$-1.67$    &4.49    &    &1.36   &$-7.55$   &2.17 \\
channel    &1.42  &$-7.53$    &2.22    &    &1.42   &$-22.06$    &1.32\\
           &1.48   &$-17.12$   &1.53     &    &1.48   &$-44.44$   &0.96\\
\bottomrule[1pt]\bottomrule[1pt]
\end{tabular}
\end{table}

We next search for possible molecular candidates mainly composed of the $KK^*$ systems. According to symmetry, the investigated quantum number configurations for the $S-$wave $KK^*$ systems include $I(J^P)=0(1^+)$ and $1(1^+)$. The discussed channels include
\begin{eqnarray*}
0(1^+): && KK^*({}^3S_1), \, KK^*({}^3D_1), \, K^*K^*({}^3S_1), \, K^*K^*({}^3D_1),\\
1(1^+): && KK^*({}^3S_1), \, KK^*({}^3D_1), \, K^*K^*({}^5D_1).
\end{eqnarray*}

In order to explore the roles of the $S-D$ wave mixing effects, the couple channel effects, and the momentum related terms in the OBE effective potentials, we present our results in several cases as shown in Table \ref{num1}, where case I and case II present the results without and with the momentum-related terms, respectively. Our analysis reveals that loosely bound state solutions for the $KK^*$ system with $I(J^P)=0(1^+)$ emerge at a cutoff parameter $\Lambda \sim 1.00$ GeV. Given that the corresponding binding energy and root-mean-square (RMS) radius satisfy the criteria for a loosely bound molecular state, we can identify the $KK^*$ system with $0(1^+)$ as a promising molecular candidate.

\begin{figure}[!htbp]
    \centering
    \includegraphics[width=0.75\linewidth]{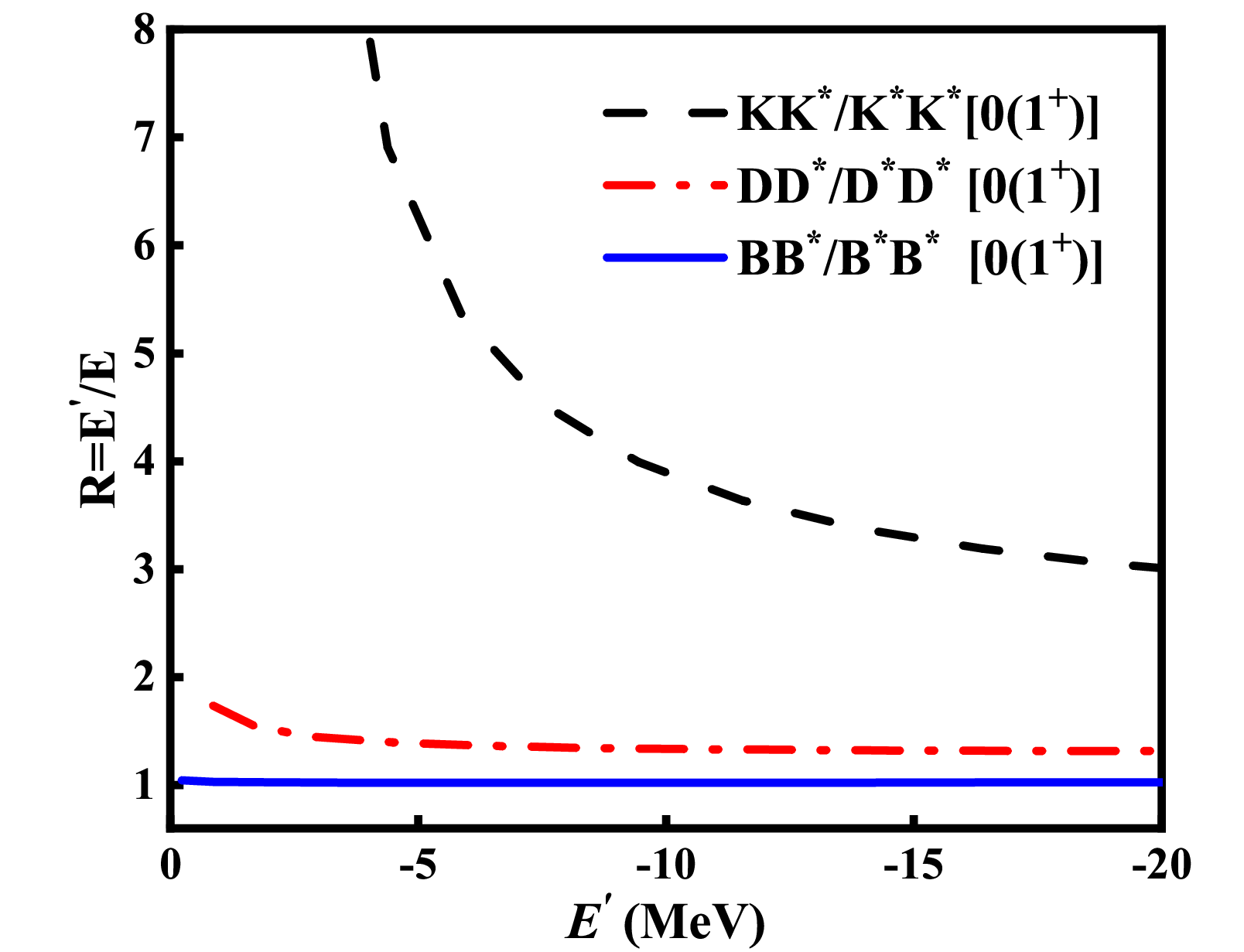}
\caption{Binding energies ratio $R=E^{\prime}/E$ for the coupled $KK^*/K^*K^*$, $DD^*/D^*D^*$, and ${B}{B}^*/{B}^*{B}^{(*)}$ bound states with $0(1^+)$. Here, $E^{\prime}$ and $E$ correspond to the binding energies with and without considering the recoil corrections, respectively.}
    \label{recoil1}
\end{figure}

Additionally, we find that: 
\begin{enumerate}
    \item The $S$-$D$ wave mixing provides an attractive contribution to the formation of this bound state, as evidenced by an increase in the binding energy compared to a pure $S$-wave analysis.
    \item The significant mass difference between the $KK^*$ and $K^*K^*$ thresholds suppresses the contributions from coupled-channel effects. This is reflected in the minimal change in the bound state properties when these channels are included.
    \item The inclusion of momentum-dependent terms in the effective potential significantly enhances the binding. As shown in Table \ref{num1} and Figure \ref{recoil1}, the binding energy increases by a factor of several for the same cutoff value. Specifically, the ratio $R = E^{\prime}/E$, where $E^{\prime}$ and $E$ denote the binding energies for Case I and Case II, respectively, can exceed 10 for extremely shallow states. This ratio gradually decreases, approaching a value of three, as the binding energy itself increases. In summary, recoil corrections, manifested through these momentum-dependent terms, play a crucial role in the formation of this bound state.
\end{enumerate}

In this work, we also investigate the role of recoil corrections in the heavy quark sector. Under the assumption that interactions involving heavy quarks are negligible, the OBE effective potentials for the $D^{(*)}D^{(*)}$ and $B^{(*)}B^{(*)}$ systems are analogous to those of the $K^{(*)}K^{(*)}$ systems with identical quantum numbers. To test this explicitly, we neglect interactions specific to the strange quark and substitute the strange meson masses with those of the corresponding heavy mesons. Upon solving the coupled-channel Schr\"{o}dinger equations, our calculations demonstrate that the recoil corrections have negligible impact on double-bottom systems. In contrast, for the charm sector, the momentum-dependent terms are found to be significant for the formation of very loosely bound molecular states. This implies that if the $T_{cc}$ observed by the LHCb Collaboration \cite{LHCb:2021vvq,LHCb:2021auc} is indeed a $DD^*$ molecule with $I(J^P)=0(1^+)$ and a very shallow binding energy, then recoil corrections likely play a crucial role in its formation.

For the remaining $1(1^+)$, the corresponding OBE effective potentials  are insufficiently attractive to support bound states.

\begin{figure}[!htbp]

    \centering
    \includegraphics[width=0.95\linewidth]{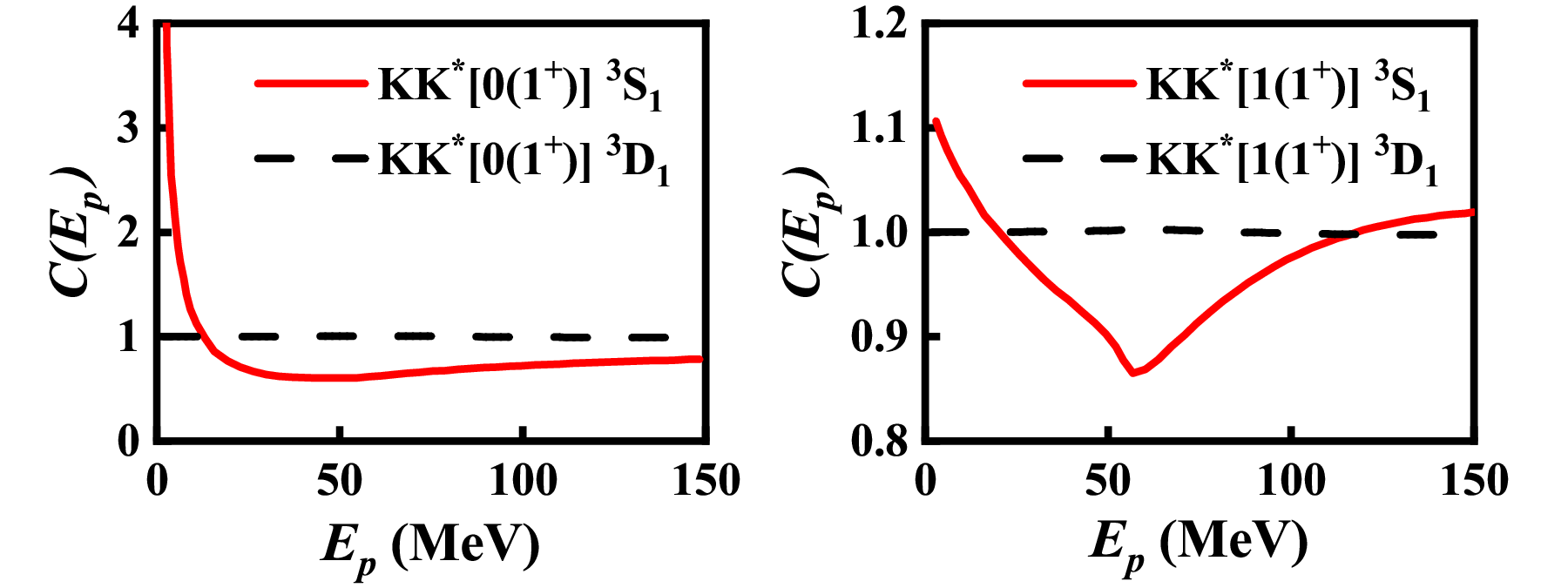}\\
    \caption{Correlation functions $C(\bm{q})$ for the $KK^*$ systems with $I(J^P)=0(1^+)$ and $1(1^+)$.}
    \label{correlation1}
\end{figure}

We then analyze the correlation functions $C(E_p)$ for the $KK^*$ systems with $I(J^P)=0(1^+)$ and $1(1^+)$. Given the minor influence of coupled-channel effects in the $KK^*/K^*K^*$ system, in Figure \ref{correlation1}, we present the correlation functions for the $KK^*$ systems with $I(J^P)=0(1^+)$ (left panel) and $1(1^+)$ (right panel) after solely considering the $S-D$ wave mixing effects. Here, the correlation functions for the $KK^*$ system with $1(1^+)$ are around 1.00, which indicates that the corresponding interactions are too week to bind $K$ and $K^\ast$ together. This finding is consistent with our direct calculation, which yielded no bound state for this channel. 

Notably, we can find that there appears a downward cusp in the correlation function of $S$-wave $1(1^+)$ system. It is because in $D$-wave interaction, there exists a repulsive barrier. After coupled to $S$-wave system, in the "interaction range" of this barrier, $K$ and $K^*$ tend to fall apart, which makes the correlation function decrease very fast. After running out of the barrier, the interaction shows mainly as a weak attraction and start to bind $K$ and $K^*$ together. As $E_p$ becomes larger and the interaction becomes weaker, the correlation function will little by little go back to one.

In contrast, for the isoscalar $S$-wave $KK^*$ system with $0(1^+)$, the appearance of a bound state makes $C(E_p)$ largest at threshold, then it decreases due to the fact that the $KK^*$ interactions with $0(1^+)$ become weak with the relative momentum becomes larger. Thus, the properties of the correlation functions confirm that the $KK^*$ system with $1(1^+)$ or $D$-wave $0(1^+)$ cannot be a good molecular candidate, while for the $KK^*$ system with $S$-wave $0(1^+)$, the correlation function satisfies the properties of the existence of a bound state.

Based on symmetry considerations, the quantum number configurations for the $S-$wave $K^*K^*$ systems can be $0(1^+)$, $1(0^+)$, and $1(2^+)$. The specific channels involved in the coupled-channel analysis are:
\begin{eqnarray*}
0(1^+): && K^*K^*({}^3S_1), \, K^*K^*({}^3D_1),\\
1(0^+): && K^*K^*({}^1S_0), \, K^*K^*({}^5D_0),\\
1(2^+): && K^*K^*({}^5S_2), \, K^*K^*({}^1D_2),\, K^*K^*({}^5D_2).
\end{eqnarray*}
In the cutoff range of $\Lambda\leq2.00$ GeV, we can only obtain the loosely bound state solutions for the $K^*K^*$ systems with $0(1^+)$. The OBE effective potentials for the remaining two systems are not strong enough to generate bound states.

\begin{table}[!htbp]
\renewcommand\tabcolsep{0.2cm}
\renewcommand{\arraystretch}{1.7}
\caption{The bound state solutions for the $K^{*}K^{*}$ system with $I(J^P)=0(1^+)$. Here, case I and case II present the results without and with the momentum-related terms, respectively. $E$ and $r_{\text{RMS}}$ stand for the binding energy and the the root-mean-square radius, respectively. The unites for the cutoff $\Lambda$, $E$, and $r_{RMS}$ are GeV, MeV, and fm, respectively.} \label{num2}
\begin{tabular}{cccccccc}
\toprule[1pt]\toprule[1pt]
    &\multicolumn{3}{c}{Case I}  &   &\multicolumn{3}{c}{Case II}\\\cline{2-4}\cline{6-8}
    &$\Lambda$    &$E$     &$r_{\text{RMS}}$ &
           &$\Lambda$    &$E$     &$r_{\text{RMS}}$\\\midrule[1pt]
$S-$wave    
&1.24  &$-0.60$  &6.14   &   &1.24   &$-3.86$   & 2.60 \\
&1.32  &$-5.85$   &2.18  &   &1.32  &$-17.85$   & 1.32 \\
&1.40  &$-16.03$   &1.41  &  &1.40  &$-43.66$ &0.89\\\hline
$S-D$   
&1.24  &$-1.34$  &4.34 &   &1.24    &$-5.37$    &2.26  \\
 wave  &1.32&$-7.69$ &1.96 & &1.32  &$-20.65$   &1.25\\
    &1.40&$-18.88$  &1.32  & &1.40  &$-47.67$   &0.88\\\hline

\bottomrule[1pt]\bottomrule[1pt]
\end{tabular}
\end{table}

\begin{figure}[!htbp]
    \centering
    \includegraphics[width=0.75\linewidth]{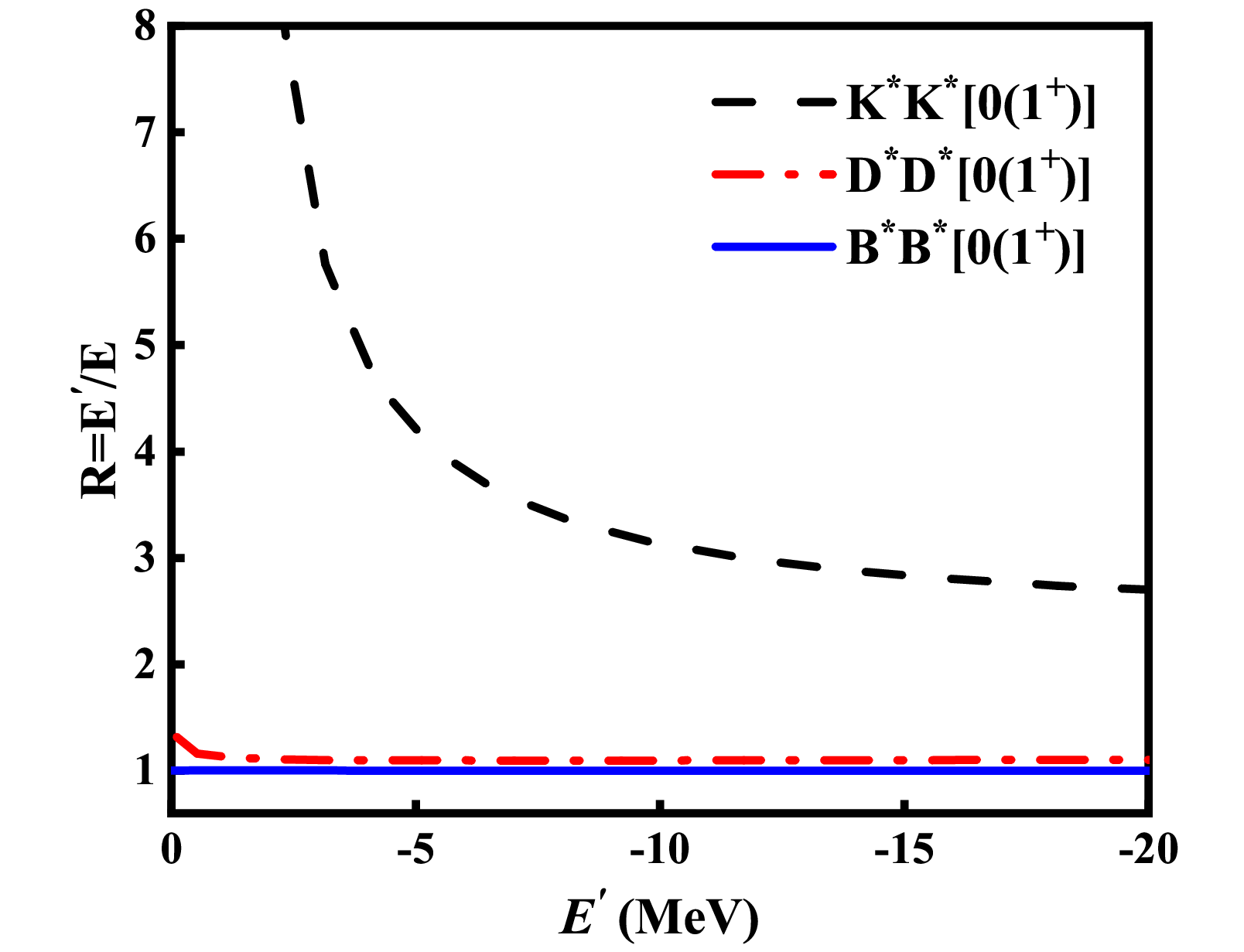}
    \caption{Binding energies ratio $R=E^{\prime}/E$ for the $K^*K^*$, $D^*D^*$, and ${B}^*{B}^{(*)}$ bound states with $0(1^+)$. Here, $E^{\prime}$ and $E$ correspond to the binding energies with and without considering the recoil corrections, respectively.}
    \label{recoil2}
\end{figure}

In Table \ref{num2}, we collect the loosely bound state solutions for the $K^{*}K^{*}$ bound state with $I(J^P)=0(1^+)$. The loosely bound state solutions emerge at the cutoff around 1.20 GeV. The corresponding binding energies and root-mean-square (RMS) radii are consistent with the properties of a loosely bound molecule, leading us to propose the $K^{*}K^{*}$ bound state with $I(J^P)=0(1^+)$ as a good molecular candidate. When comparing the $S$-wave, for the same cutoff, the binding energy can be a little larger in the $S-D$ wave mixing case. Thus, the $S-D$ wave mixing effects have a positive impact on this bound state. Furthermore, our results reaffirm the significance of recoil corrections for systems involving light mesons. As shown in Figure \ref{recoil2}, the influence of these corrections becomes increasingly pronounced with decreasing meson mass. For binding energies smaller than approximately $-10$ MeV, the ratio $R = E^{\prime}/E$ of the binding energies with ($E^{\prime}$) and without ($E$) recoil corrections can exceed a factor of 3. While this ratio stabilizes for more deeply bound states, recoil corrections remain non-negligible in the strange sector. In contrast, for systems in the heavy-flavor sector, the contributions from recoil corrections are found to be minimal.

Similarly as our previous discussions on $KK^*$ system, for $K^*K^*$ system with $I(J^P)=0(1^+)$ (lower-left panel), since there exists bound state solutions as given in Table~\ref{num2}, the correlation function of this $0(1^+)~K^*K^*$ behaves just the same as the left panel of Fig.~\ref{correlation1}. While for $1(0^+)$ (upper-left panel) and $1(2^+)$ (upper-left panel) configurations, since the main characters of the interactions are repulsive, the $K^*K^*$ tend to fall apart at this time, especially at $E_p=0$ (MeV), where the repulsive effect is largest. Thus, the correlation functions start from a value below one, then, with the increase of relative energy and decrease of interaction, the correlation function get close to one.

\begin{figure}[!htbp]
    \centering
       \includegraphics[width=0.95\linewidth]{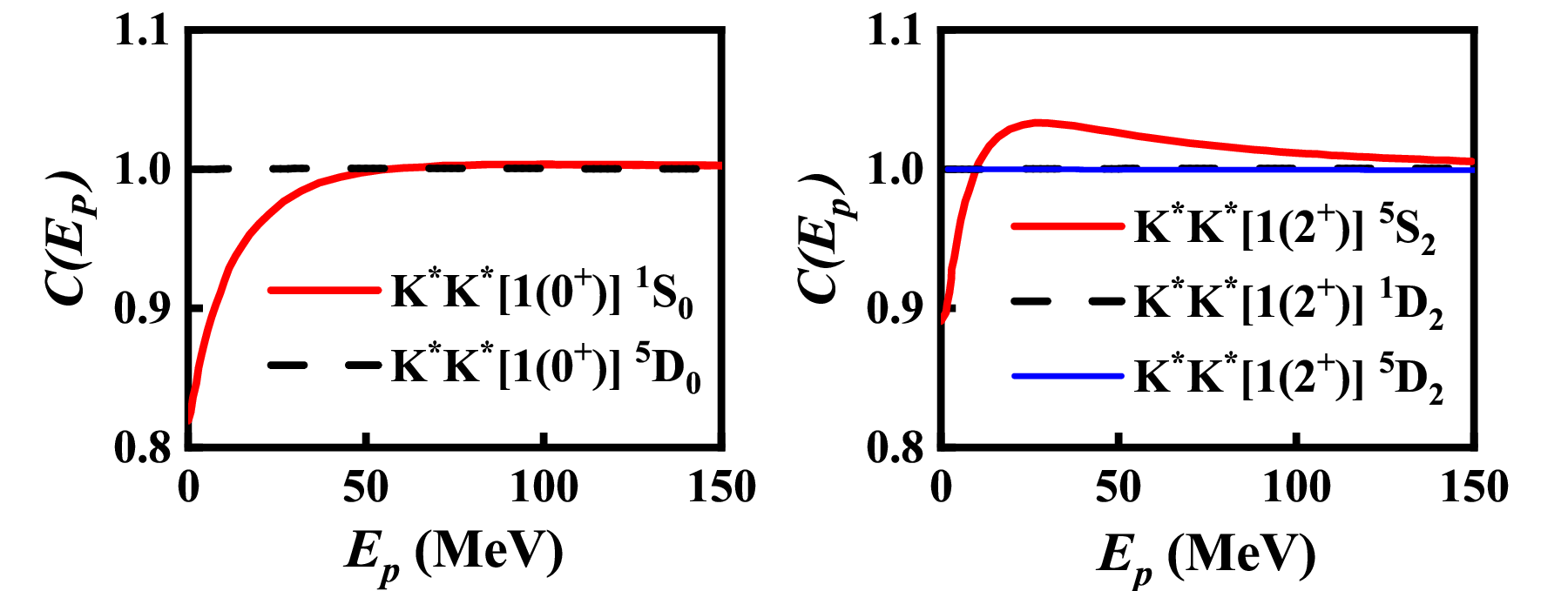} 
       \includegraphics[width=0.96\linewidth]{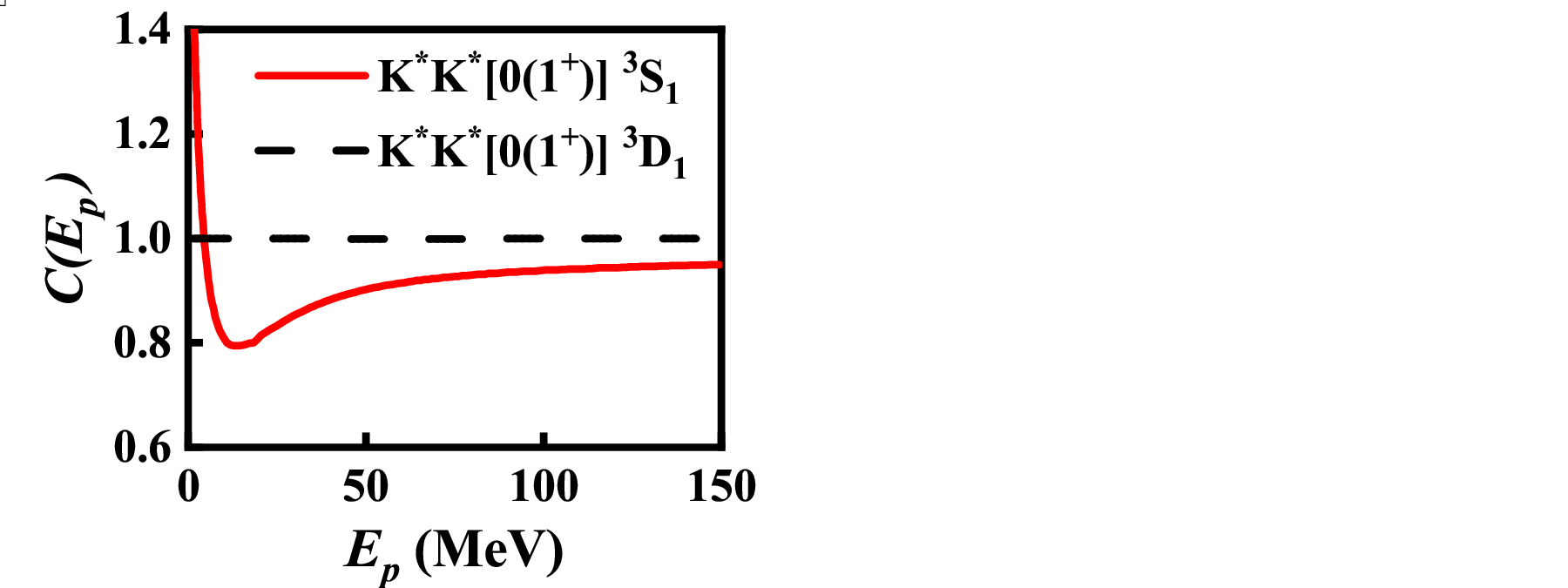}
    \caption{Correlation function for the $K^*K^*$ system with $I(J^P)=0(1^+)$, $1(0^+)$ and $1(2^+)$.}
    \label{recoil3}
\end{figure}

In summary, we can predict two possible molecular candidates, the $KK^*$ molecule with $0(1^+)$ and the $K^*K^*$ molecule with $0(1^+)$. Our analysis demonstrates that recoil corrections are crucial for the formation of these loosely bound states. Furthermore, the $S-D$ wave mixing effects play a positive role.  

\subsection{The $K^{(*)}\bar{K}^{(*)}$ systems}

As a byproduct, we extend to study the $K^{(*)}\bar{K}^{(*)}$ systems within the framework of the OBE model. Here, we also consider the $S-D$ wave mixing effects. The quantum number configurations of the $K\bar{K}^*$ and $K^*\bar{K}^*$ systems can be
\begin{eqnarray}\begin{array}{ccccc}
K\bar{K}^*(I|{}^{2S+1}L_J\rangle) :  &|{}^3S_1\rangle,&|{}^3D_1\rangle,\\
K^*\bar{K}^*(I|{}^{2S+1}L_J\rangle) :  &|{}^1S_0\rangle,&|{}^5D_0\rangle,\\
                                      &|{}^3S_1\rangle, &|{}^3D_1\rangle,&|{}^5D_1\rangle,\\
                                      &|{}^5S_2\rangle, &|{}^1D_2\rangle,&|{}^3D_2\rangle,&|{}^5D_2\rangle.
\end{array}\end{eqnarray}
The flavor wave functions for the $K\bar{K}^*$ systems can be constructed as
\begin{eqnarray*}
|1,1\rangle &=& \frac{1}{\sqrt{2}}\left(|K^+\bar{K}^{*0}\rangle
                +c|K^{*+}\bar{K}^0\rangle\right),\\
|1,-1\rangle &=& \frac{1}{\sqrt{2}}\left(|K^0{K}^{*-}\rangle
                  +c|K^{*0}{K}^-\rangle\right),\\
|1,0\rangle &=& \frac{1}{{2}}\left(|K^+{K}^{*-}\rangle-|K^0\bar{K}^{*0}\rangle
              +c|K^{*+}{K}^{-}\rangle-c|K^{*0}\bar{K}^{0}\rangle\right),\nonumber\\
|0,0\rangle &=& \frac{1}{{2}}\left(|K^+{K}^{*-}\rangle+|K^0\bar{K}^{*0}\rangle
              +c|K^{*+}{K}^{-}\rangle +c|K^{*0}\bar{K}^{0}\rangle\right).
\end{eqnarray*} 
Here, $c=\pm$ corresponds to C-parity $C=\mp$, respectively. The flavor wave functions for the $K^*\bar{K}^*$ systems can be constructed as
\begin{eqnarray}
|1,1\rangle &=& \left(|K^{*+}\bar{K}^{*0}\rangle\right),\\
|1,-1\rangle &=& \left(|K^{*-}{K}^{*0}\rangle\right),\\
|1,0\rangle &=& \frac{1}{\sqrt{2}}\left(|K^{*+}{K}^{*-}\rangle
                 -|K^{*0}\bar{K}^{*0}\rangle\right),\\
|0,0\rangle &=& \frac{1}{\sqrt{2}}\left(|K^{*+}{K}^{*-}\rangle
                 +|K^{*0}\bar{K}^{*0}\rangle\right).
\end{eqnarray}

According to the $G-$parity rule, we can obtain the relations for the OBE effective potentials between the $K^{(*)}\bar{K}^{(*)}$ systems and $K^{(*)}{K}^{(*)}$ systems, i.e.,
\begin{eqnarray}
V_{M}^{K^{(*)}\bar{K}^{(*)}\to K^{(*)}\bar{K}^{(*)}}=(-1)^{G_M}V_{M}^{K^{(*)}{K}^{(*)}\to {K^{(*)}{K}^{(*)}}},
\end{eqnarray}
where $M$ labels as the exchanged mesons, and $G_M$ is the $G-$parity for the exchanged mesons.

\begin{table}[!htbp]
\renewcommand\tabcolsep{0.15cm}
\renewcommand{\arraystretch}{1.7}
\caption{The bound state solutions for the $K^{(*)}\bar{K}^{*}$ systems. Here, case I and case II present the results without and with the momentum-related terms, respectively. $E$ and $r_{\text{RMS}}$ stand for the binding energy and the the root-mean-square radius, respectively. The unites for the cutoff $\Lambda$, $E$, and $r_{RMS}$ are GeV, MeV, and fm, respectively.} \label{num3}
\begin{tabular}{cccccccc}
\toprule[1pt]\toprule[1pt]
    &\multicolumn{3}{c}{Case I}  &   &\multicolumn{3}{c}{Case II}\\\cline{2-4}\cline{6-8}
    &$\Lambda$    &$E$     &$r_{\text{RMS}}$ &
           &$\Lambda$    &$E$     &$r_{\text{RMS}}$\\\midrule[1pt]
    
$K\bar{K}^*[0^{-}(1^{+-})]$       
           &2.00  &$-2.50$  &3.89  &    &1.53   &$-0.96$   &5.83\\
$S-$wave   &2.20  &$-10.35$ &2.04   &    &1.54  &$-3.70$   & 3.09 \\
          &2.40  &$-20.04$  &1.45 &    &1.55  &$-29.17$   &1.05\\\hline
 
 $S-D$      &1.60  &$-0.44$   &8.58 &   &1.53   &$-8.20$    &2.18  \\
 wave&1.80  &$-10.61$  &2.05  &  &1.54  &$-16.28$    &1.57\\
         &2.00  &$-37.45$  &1.23 &   &1.55  &$-33.21$    &1.07\\\midrule[1.5pt]
  $K\bar{K}^*[0^{+}(1^{++})]$       
           &1.70  &$-1.36$  &5.03  &    &1.43   &$-0.37$   &8.39\\
$S-$wave   &1.90  &$-9.36$ &2.10   &    &1.49  &$-6.57$   & 2.32 \\
          &2.10  &$-20.60$  &1.48 &    &1.55  &$-25.43$   &1.22\\\hline
 
$S-D$        &1.70  &$-9.08$   &2.14 &   &1.43   &$-2.60$    &3.64  \\
 wave                             &1.90  &$-25.28$  &1.37  &  &1.49  &$-11.95$    &1.79\\
                &2.10  &$-44.55$  &1.09 &   &1.55  &$-33.69$    &1.09\\\midrule[1.5pt]
$K^*\bar{K}^*[0^{+}(0^{++})]$       
           &1.04  &$-0.13$  &10.29 &    &1.04   &$-0.84$   &5.25\\
$S-$wave   &1.10  &$-7.45$ &1.95   &    &1.10  &$-13.70$   & 1.48\\
          &1.16  &$-27.25$  &1.11 &    &1.16  &$-47.25$   &0.87\\\hline
 
 $S-D$         &1.04  &$-0.43$   &6.98 &   &1.04   &$-1.42$    &4.16  \\
  wave  &1.10  &$-8.75$  &1.83  &  &1.10  &$-15.33$    &1.41\\
     &1.16  &$-29.30$  &1.09 &   &1.16 &$-49.69$    &0.86\\\midrule[1.5pt]
 $K^*\bar{K}^*[0^{-}(1^{+-})]$      &1.18 &$-0.02$    &13.00 & & 1.18  &$-1.49$   &4.09 \\
       $S-$wave                  &1.24  &$-3.73$   &2.66 & & 1.24  &$-13.09$    &1.51\\
                           &1.30  &$-13.66$   &1.52  & &1.30   &$-39.70$   &0.94\\\hline
 $S-D$      &1.18  &$-0.28$    &8.24  & &1.18   &$-2.42$   &3.25 \\
   wave &1.24  &$-5.01$   &2.35  & &1.24  &$-15.14$    &1.43\\
    &1.30  &$-15.69$   &1.44  & &1.30   &$-42.56$   &0.92\\\midrule[1.5pt]
  $K^*\bar{K}^*[0^{+}(2^{++})]$       
           &1.84  &$-0.01$  &13.75  &    &1.84   &$-0.56$   &6.51\\
$S-$wave   &2.00  &$-1.00$ &5.14   &    &2.00  &$-6.56$   & 2.20\\
          &2.16  &$-3.32$  &2.98 &    &2.16  &$-39.29$   &1.04\\\hline
 
 $S-D$       &1.84  &$-0.17$   &9.83&   &1.84   &$-1.15$    &4.81  \\
 wave            &2.00  &$-1.74$  &4.02 &  &2.00 &$-8.88$    &1.95\\
      &2.16  &$-4.84$  &2.55 &   &2.16  &$-48.47$    &0.96\\
\bottomrule[1pt]\bottomrule[1pt]
\end{tabular}
\end{table}

In Table \ref{num3}, we present the bound state solutions for the $K^{(*)}\bar{K}^*$ systems. Our results show that: 
\begin{itemize}
    \item We can obtain the loosely bound state solutions for all the isoscalar states with the cutoff $\Lambda\leq2.5$ GeV. However, for the isovector systems, there cannot exist loosely bound state solutions in the same cutoff region, thus, in this case the OBE effective potentials cannot provide strong enough attractive interactions. 
    \item For the $K\bar{K}^*$ states with $0^-(1^{+-})$ and $0^+(1^{++})$, the $K\bar{K}^*$ states with $0^+(1^{++})$ binds a little bit deeper by taking the same cutoff values. 
    \item For the isoscalar $K^*\bar{K}^*$ bound states, we obtain a cutoff relation, $\Lambda(0^{++})<\Lambda(1^{+-})<\Lambda(2^{++})$, which indicates the OBE effective potentials for the $K^*\bar{K}^*$ state with $0^+(0^{++})$ is strongest attractive, followed by the $0^-(1^{+-})$ and $0^+(2^{++})$. This finding is consistent with our previous results for the $D^{*}\bar{D}^{*}$ system \cite{Wang:2021aql}.
    \item The inclusion of $S-D$ wave mixing effects yields a moderate increase in the binding energies, confirming that these effects provide an additional attractive contribution to the binding.
    \item After considering the momentum-related terms, we can obtain the loosely bound state solutions with a smaller cutoff values than those in Case I, which demonstrate that the recoil corrections play a crucial role in the formation of these bound states.
\end{itemize}

Based on the obtained bound state solutions within a reasonable range of cutoff parameters, we conclude that the $K\bar{K}^*$ states with $0^-(1^{+-})$ and $0^+(1^{++})$, and the $K^*\bar{K}^*$ states with $0^+(0^{++})$, $0^-(1^{+-})$, and $0^+(2^{++})$ can be suggested as good molecular candidates. Our findings are consistent with previous explorations of $K^{(*)}\bar{K}^*$ molecules. For example, in Ref. \cite{Lu:2016nlp}, authors studied the $K\bar{K}^*$ interactions in a quasipotential Bethe-Salpeter equation approach combined
 with the one-boson-exchange model, and they also obtained the $K^{(*)}\bar{K}^*$ bound states with $0^+(1^{++})$ and $0^-(1^{+-})$, which can related to the $f_1(1285)$ and $h_1(1380)$, respectively.

\begin{figure}[!htbp]
    \centering
    \includegraphics[width=0.95\linewidth]{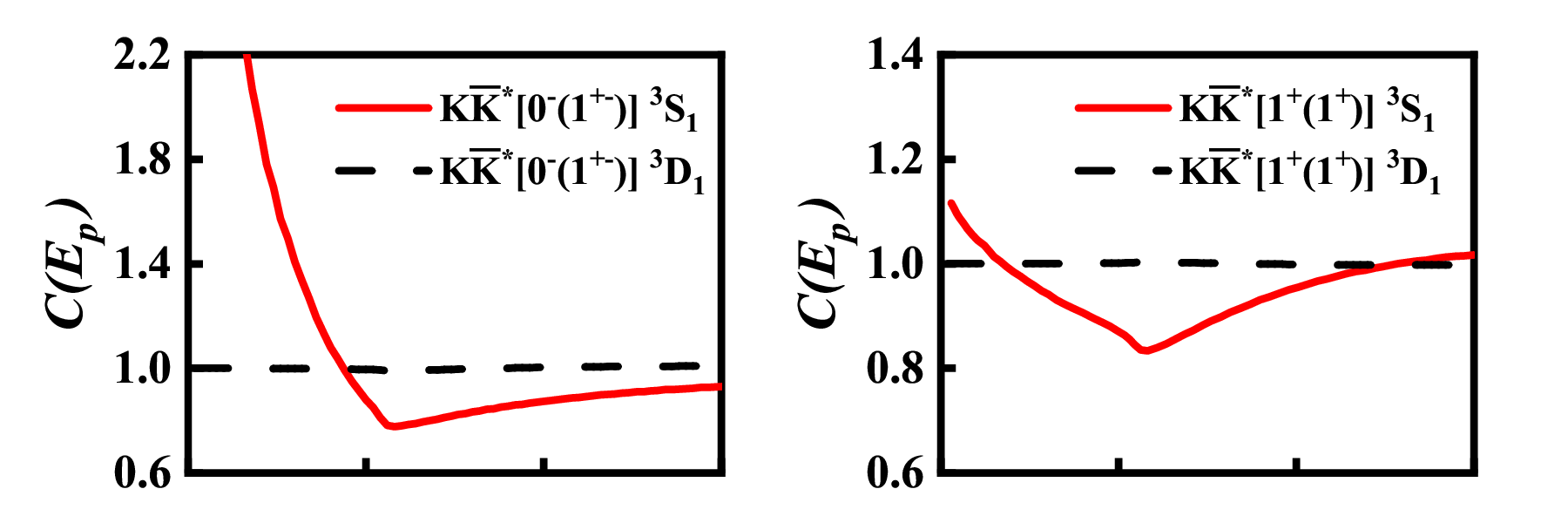}
    \includegraphics[width=0.95\linewidth]{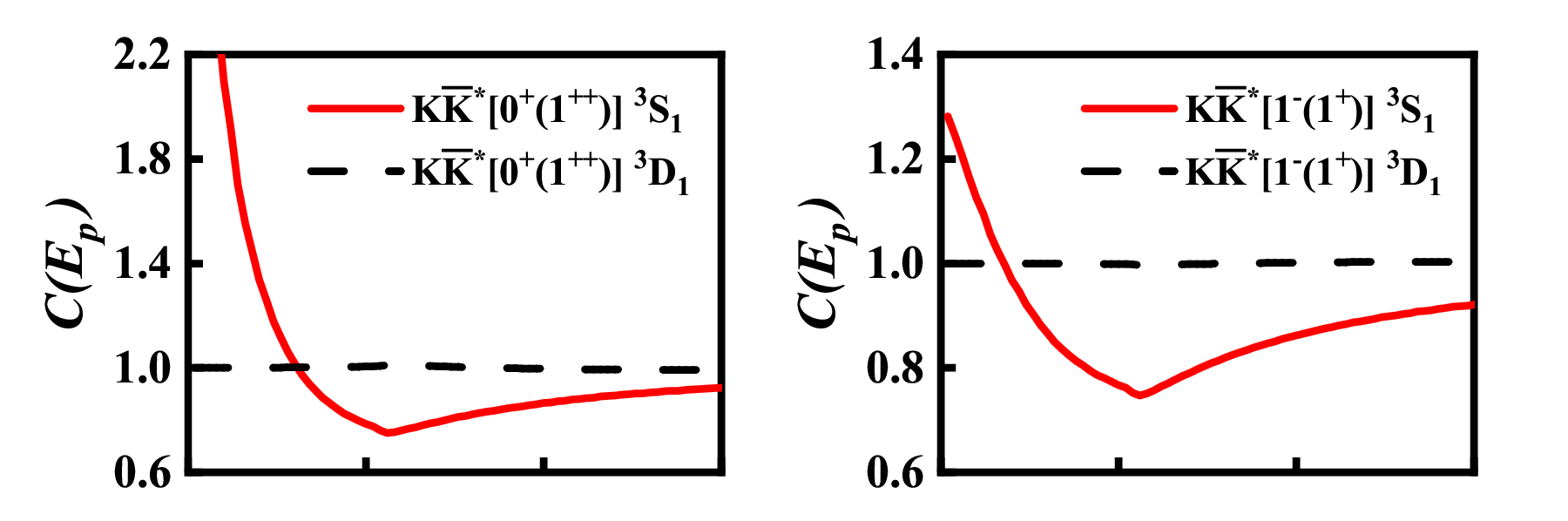}
    \includegraphics[width=0.95\linewidth]{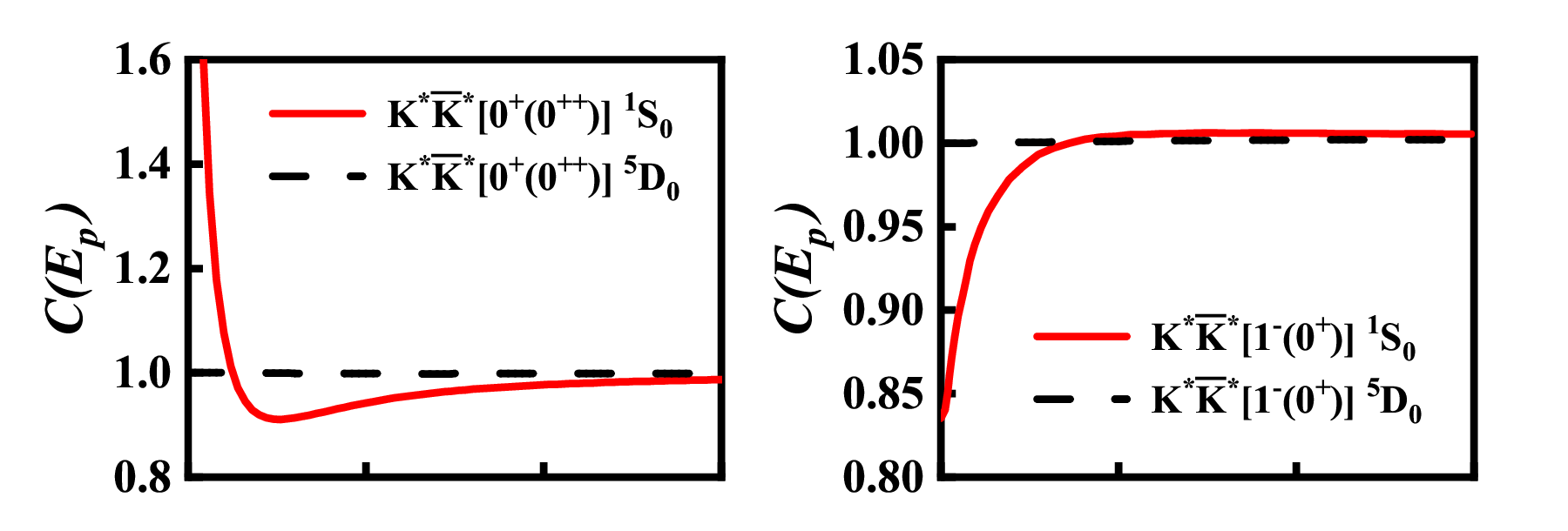}\\ \includegraphics[width=0.95\linewidth]{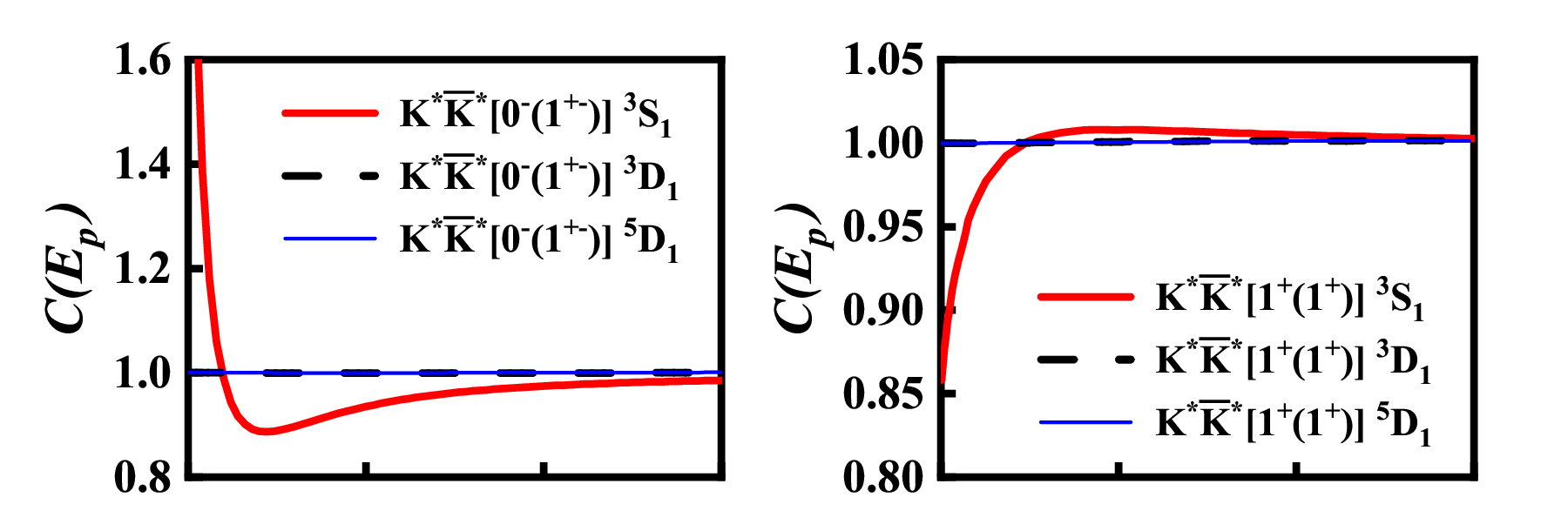}\\ \includegraphics[width=0.95\linewidth]{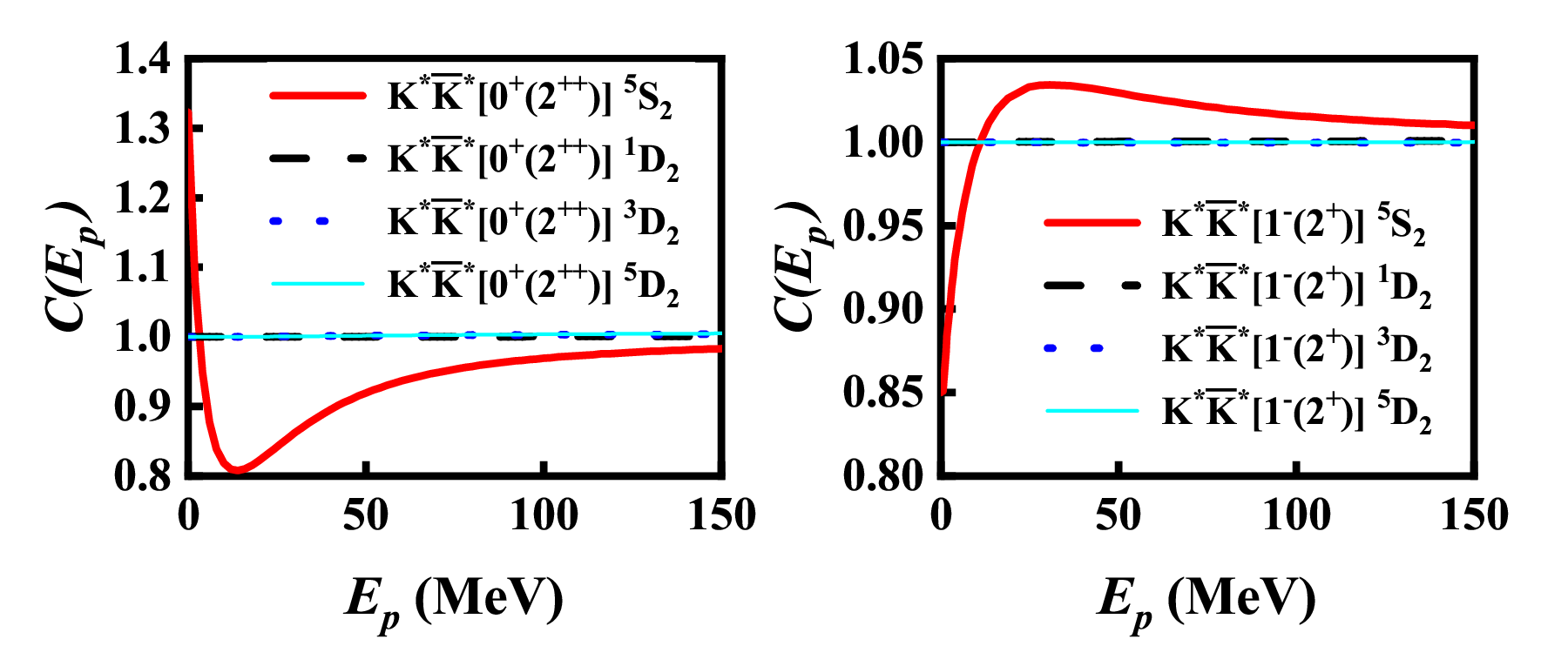}\\
    \caption{Correlation functions for all the discussed $K^{(*)}\bar{K}^*$ systems.}
    \label{recoil4}
\end{figure}

In Fig.~\ref{recoil4}, we present the correlation functions of the $K^{(*)}\bar{K}^*$ systems. From the figures on the left side, we can easily see that there do exist isoscalar $K^{(*)}\bar{K}^*$ bound states, since their behaviors are just the same as those of isoscalar $K^{(*)}K^*$ systems. In addition, from the value of $C(E_p=0)$, we can eaily infer that the stronger the interaction, the bigger the correlation function at $C(E_p=0)$. Such phenomena behaves very explicitly in the correlation functions of isoscalar $K^*\bar{K}^*$ system, it is because at $E_p=0$, the correlation function has the biggest value for $0^+(0^{++})$ configuration, followed by the $0^-(1^{+-})$ and $0^+(2^{++})$. For the isovector $K\bar{K}^*$ systems (right side of Fig.~\ref{recoil4}), the correlation functions show that although there exists weak interactions there, they can not overcome the effect of repulsive barrier, so there are no bound state solutions. For the isovector $K^*\bar{K}^*$ systems, the main character of interaction is totally repulsive, so the correlation functions will start below one, which also makes bound state solutions impossible.

\section{Summary}\label{sec4}

In the past decades, the experiments have discovered a variety of rich hadron structures, which has provided us with an excellent opportunity to study hadron spectra and the interactions between hadrons. The continuous accumulation of experimental data and the improvement of experimental precision not only bring new opportunities for studying the hadron spectrum, but also impose higher demands on the accuracy of theoretical models.

In this work, we have conducted a systematic investigation of the $K^{(*)}K^{(*)}$ interactions by using the OBE model. Our derivation of the effective potentials incorporates $S-D$ wave mixing and coupled-channel effects, with terms retained up to $\mathcal{O}(1/M^2)$. This approach comprehensively includes spin-spin interactions, spin-orbit couplings, tensor forces, and recoil corrections.

By numerically solving the coupled-channel Schrödinger equation with these potentials, we can identify two promising double-strange molecular candidates: the $KK^*$ molecule with $I(J^P) = 0(1^+)$ and the $K^*K^*$ molecule with $0(1^+)$. Our analysis demonstrates that recoil corrections are essential for the formation of these molecular states. While the $S-D$ wave mixing effects can provide additional attractive contributions, the coupled-channel effects are found to have a negligible impact on the binding.

As a supplementary study, we examined the role of recoil corrections in the heavy quark sector. We find that these corrections are non-negligible for loosely bound charm-sector molecules but have a minimal effect in the bottom sector due to the larger meson masses.

We further extend our framework to study the $K^{(*)}\bar{K}^{(*)}$ interactions. Their OBE effective potentials can relate to those in the $K^{(*)}{K}^{(*)}$ systems in the $G-$parity rule. Our calculations predict several viable strangonium-like molecular candidates: the $K\bar{K}^{*}$ molecules with $0(1^{+-}, 1^{++})$, the $K^{*}\bar{K}^{*}$ molecules with $0(0^{++}, 1^{+-}, 2^{++})$. In contrast, the OBE potentials for isovector systems lack sufficient strength to generate bound states. The importance of recoil corrections and $S-D$ wave mixing is reaffirmed in these systems.

Additionally, we compute the momentum correlation functions for the interacting meson pairs. The resulting correlations can provide independent support for strong attractive interactions in the $K^{(*)}K^*$ systems with $0(1^+)$, the $K\bar{K}^*$ systems with $0(1^{+-}, 1^{++})$ and the $K^{(*)}\bar{K}^*$ systems with $0(0^{++}, 1^{+-}, 2^{++})$. Conversely, the correlation functions for other configurations indicate weakly attractive or repulsive interactions, consistent with the absence of bound states in those channels.

Our specific predictions for the double strange and strangonium-like molecular states can be tested in experiments, particularly through amplitude analyses of processes such as $\Upsilon$ decays at Belle II or in proton-proton collisions at LHCb. The calculated correlation functions also provide a benchmark for future femtoscopic studies in heavy-ion collisions.

\section*{Acknowledgement}
This work is supported by the National Natural Science Foundation of China under Grants Nos. 12305139, 12305087. R. C. is supported by the Xiaoxiang Scholars Programme of Hunan Normal University. Q. H. is supported the Start-up Funds of Nanjing Normal University under Grant No.~184080H201B20.



\end{document}